\documentclass[%
 reprint,
 amsmath,amssymb,
 aps,
prb,
]{revtex4-2}
\usepackage{graphicx}
\usepackage{dcolumn}
\usepackage{bm}

\usepackage{xcolor}
\usepackage{epstopdf}
\usepackage{rotating}
\usepackage{natbib}
\usepackage[section]{placeins}
\usepackage{mdwlist}
\usepackage{lipsum}
\usepackage{hyperref}
\usepackage[normalem]{ulem}

\usepackage[caption=false, position=t,singlelinecheck=off]{subfig}

\usepackage{tabularx}

\newcommand {\be}{\begin{equation}}
\newcommand {\ee}{\end{equation}}
\newcommand{\ket}[1]{|#1\rangle}

\newcommand{\expect}[1]{\langle#1\rangle}
\newcommand{\braket}[2]{\langle#1|#2\rangle}
\newcommand{\ketbra}[2]{|#1\rlap{/}\backslash#2|}
\newcommand{\varg}{ \text{\it{g}} }

\begin{document}


\title{Reservoir Computing Approach to Quantum State Measurement}
\author{Gerasimos Angelatos,
Saeed Khan,
Hakan E.~T\"{u}reci}

\affiliation{Department of Electrical Engineering, Princeton University, Princeton, New Jersey 08544, USA}

             
\begin{abstract}
Efficient quantum state measurement is important for maximizing the extracted information from a quantum system. For multi-qubit quantum processors in particular, the development of a scalable architecture for rapid and high-fidelity readout remains a critical unresolved problem.  Here we propose reservoir computing as a resource-efficient solution to quantum measurement of superconducting multi-qubit systems. We consider a small network of Josephson parametric oscillators, which can be implemented with minimal device overhead and in the same platform as the measured quantum system.  We theoretically analyze the operation of this Kerr network as a reservoir computer to classify stochastic time-dependent signals subject to quantum statistical features. We apply this reservoir computer to the task of multinomial classification of measurement trajectories from joint multi-qubit readout. For a two-qubit dispersive measurement under realistic conditions we demonstrate a classification fidelity reliably exceeding that of an optimal linear filter using only two to five reservoir nodes, while simultaneously requiring far less calibration data \textendash{} as little as a single measurement per state.  
We understand this remarkable performance through an analysis of the network dynamics and develop an intuitive picture of reservoir processing generally. Finally, we demonstrate how to operate this device to perform two-qubit state tomography and continuous parity monitoring with equal effectiveness and ease of calibration.  This reservoir processor avoids computationally intensive training common to other deep learning frameworks and can be implemented as an integrated cryogenic superconducting device for low-latency processing of quantum signals on the computational edge.  
\end{abstract}
\maketitle



Rapid and high-fidelity single-shot readout is an important primitive for manipulation and processing of quantum information. In superconducting circuit quantum processors~\cite{Gambetta2017, Kjaergaard2020}, this requires a careful calibration of the entire measurement chain, including cryogenic and room-temperature amplifiers, circulators, attenuators and room-temperature electronics. This calibration becomes particularly resource-intensive for readout systems attached to multi-qubit quantum processors. The optimization of quantum state readout has therefore been the focus of considerable ongoing research \cite{Gambetta2007,Gambetta2008, Filipp2009, Lalumiere2010, Jeffrey2014, Heinsoo2018, Ikonen2019, Boulant2007, Boissonneault2010, Reed2010}, involving a delicate balance of competing requirements: fidelity and speed.

For single-qubit readout, optimal filtering approaches~\cite{Gambetta2007} and hardware architectures have been developed and implemented to achieve fast and high-fidelity measurements without affecting qubit coherence~\cite{Walter2017}. More recently, recognizing that the quantum measurement problem in its very essence is the classification of time-dependent voltage signals acquired at the end of a measurement chain, machine learning solutions have been investigated~\cite{Magesan2015,Flurin2020,Palmieri2020}, and have shown an increase in single-qubit state discrimination by a few percent with respect to these conventional approaches~\cite{Magesan2015}. For measurement in multi-qubit systems however, the optimization and calibration of a readout system presents a difficult hardware design as well as a computationally intensive signal processing problem \cite{Filipp2009, Jeffrey2014, Heinsoo2018}; measurement cross-talk in particular imposes significant limitations on device scaling~\cite{Heinsoo2018}.  Here we propose the integration of reservoir computing as a novel hardware-efficient approach to high-fidelity multi-qubit state readout and its training-based calibration. 

Reservoir computing is a machine learning framework for the processing of time-dependant data~\cite{Jaeger2004, Jaeger2009, Sande2017, Tanaka2019}.  It is founded on the idea that any sufficiently complex and high-dimensional dynamical system, where only the linear output layer is optimized, can have the same computational capacity as a recurrent neural network and approximate arbitrary functions~\cite{Dambre2012, Gonon2021}. 
Vastly different physical systems have been employed as Reservoir Computers (RCs) for applications such as forecasting and classification~\cite{Appeltant2011,Haynes2015,Larger2017,Du2017,Coulombe2017,Canaday2018,Griffith2019}. The field of reservoir computing has recently expanded to include quantum systems~\cite{Fujii2017, Ghosh2019, Schuld2019, Govia2020, Ghosh2020, Chen2020}.  However, the application of reservoir processing to the problem of quantum measurement has not yet been explored.

Our goal in this paper is three-fold: (1) Describe a reservoir computing approach to quantum measurement that utilizes a physical system with recurrent connections, (2) analyze its efficacy for fast and high-fidelity readout {and monitoring} of multiple qubits simultaneously, (3) propose a superconducting pre-processor based on a network of Josephson Parametric Oscillators to enable hardware efficient and low latency multi-qubit measurement. While we discuss this approach for a multi-qubit superconducting platform, and a corresponding Josephson junction-based superconducting reservoir, we anticipate that the techniques are general enough to also be applicable for a broader class of quantum systems, measurement tasks, and reservoirs.

Conventional RC wisdom suggests that very high-dimensional dynamical systems are necessary for strong computational performance, with $10^2$-$10^3$ nodes typically being used \cite{Appeltant2011,Pathak2018} in software or time-delay architectures (where there is less overhead associated with increasing the size of the network). 
Here, we show that a small physical RC (2-5 nodes) is able to classify two-qubit measurement trajectories with a fidelity that is higher than achievable under the same conditions with conventional optimal filtering approaches.
{Equally strong performance is seen across a variety of quantum systems and measurement tasks, without requiring any modification of the RC.
}
This non\textendash{}von-Neumann\textendash{}architecture computer can be implemented in the same hardware platform as the target quantum system with minimal overhead, providing a uniquely low latency approach to quantum measurement.
An important conclusion of our study is that the Kerr network RC we consider learns significantly faster than a readout system that is calibrated using an optimal matched filter. Our results indicate that a cryogenic readout device should provide a rapid, robust and autonomous pre-processor for quantum state measurement. Such analog processors are capable of operating on timescales orders of magnitude faster than digital processors in head-to-head comparison on the same computational task \cite{Canaday2020}, and enable signal processing on the `computational edge,' \cite{Canaday2018, Tanaka2019} significantly reducing computational costs.

In addition, we demonstrate that the RC provides a model-independent approach to quantum state measurement, ideal for multi-qubit systems whose readout chains are projected to become increasingly more complex. The readout problem we consider here is one of retrodicting certain features of the initial state of a measured Quantum System (QS), based on information obtained after a specific quantum process, such as the scattering of a probe pulse off of the QS. Within superconducting circuit implementations, the most widely-employed readout setup is that of quantum non-demolition (QND) dispersive measurement~\cite{Gambetta2007, Gambetta2008, Filipp2009, Jeffrey2014, Heinsoo2018, Ikonen2019}; however, actual hardware implementations exhibit non-QND effects and experimental imperfections such as drift and cross-talk. Such non-idealities are difficult to optimize in hardware and require several calibration experiments to characterize, making precise knowledge of the implemented physical model difficult to acquire. This lack of an accurate physical model generally rules out a description of the measurement chain via a stochastic master equation (SME), integration of which would predict \textit{precisely} the measurement signal obtained given any initial state of the QS.

The difficulty of extracting the implemented physical model increases the appeal of model-independent approaches, such as linear filtering of the experimental data (discussed in Sec~\ref{sec:quantumsystem}); however this typically requires a large amount of training data and is susceptible to errors from quantum jumps and qubit decay. An alternative model-independent machine learning approach has been taken in Ref.~\onlinecite{Flurin2020}, although a practical application to multi-qubit measurement has yet to be demonstrated, and will likely be limited by computing capacity. In this paper we apply an RC to a scenario in which constraints on QNDness and cross-talk in multi-qubit readout are relaxed, thus simulating quantum measurement with unoptimized hardware where the complexity of readout is relegated to the processing of acquired signals. We find that the RC is able to perform said processing with high fidelity and minimal computational cost, enabling a powerful model-free approach to readout.  Specifically, we consider a situation where two qubits are measured simultaneously through a common resonator, without dedicated readout cavities and Purcell filters; our objective is not to propose this particular measurement scheme, but rather emphasize the reduced hardware and optimization overhead, and thus increased potential scalability, enabled through our reservoir processing approach generally.

We begin in Sec.~\ref{sec:quantumsystem} with a description of the joint dispersive readout task we consider in this work.
We then give a high-level overview of our proposed RC quantum measurement system in Sec.~\ref{subsec:proposal}, followed by a detailed description of the Kerr network RC model in Sec.~\ref{subsec:KerrRCmodel}
Typical dynamics and performance of a specific Kerr RC classifying quantum measurement records are presented in Sec.~\ref{subsec:RCdyn}.  We then demonstrate the ability of a Kerr RC to rapidly learn a quantum measurement task in Sec.~\ref{subsec:rapidcal}, enabling fast readout calibration. In Sec.~\ref{subsec:phasedyn}, we use an analysis of the Kerr RC phase space dynamics to explain the strong performance of RCs with as few as two Kerr nodes and develop an intuitive picture of RC processing.  In Sec.~\ref{subsec:RCdesign} we explore how behaviour varies with system hyperparameters and present basic principles for the optimization of a hardware RC. 
{Finally, in Sec.~\ref{sec:Qinfo} we demonstrate how one can operate this reservoir processor to perform two additional important quantum information tasks: two-qubit state tomography and continuous parity monitoring.
}

\section{Quantum state Readout of Multiple Qubits}
\label{sec:quantumsystem}


Multi-qubit readout presents a sufficiently difficult problem to quantitatively {assess} the advantage provided by more sophisticated signal processing techniques. We therefore discuss this problem in some detail below, {leaving some of the mathematical details for Appendix~\ref{app:readout}}. Extensions of single-qubit quantum readout approaches to larger multi-qubit systems through various multiplexing techniques have been extensively investigated \cite{Filipp2009,Lalumiere2010, Jeffrey2014,Heinsoo2018}. A majority of these schemes rely on the premise of quantum non-demolition (QND) measurement through the dispersive readout technique. In the single-qubit variant, the binary state of the qubit ($\ket{0}$, $\ket{1}$) is encoded in the amplitude and phase of a microwave pulse scattered from a readout resonator that is dispersively coupled to the qubit \cite{Gambetta2008}.

A significant problem {that scales very unfavourably with system size in multi-qubit devices is cross-talk}, which we take here to be any effect of the measurement process on parts of the system one is not trying to measure. Reducing readout errors generally requires precise calibration of readout pulses, carefully designed Purcell filters and a chip layout that minimizes cross-talk~\cite{Jeffrey2014, Heinsoo2018}. Such optimized calibration is difficult for a practical multi-qubit quantum processor due to drifts in system parameters, and reducing cross-talk imposes severe limitations on readout spectral bandwidth. Here we consider the joint dispersive readout scenario~\cite{Filipp2009, Lalumiere2010} where all qubits are coupled to the same mode of a common readout resonator, {with the goal of measuring} the combined state of all qubits in {a single} shot.

Our starting point is the multi-qubit Jaynes-Cummings (JC) Hamiltonian:
\begin{align}
\hat{\mathcal{H}}_{JC} = &\Delta_c \hat{d}^{\dagger}\hat{d}  +\epsilon(t)(\hat{d}+\hat{d}^\dagger)
 \nonumber \\
&+ \sum_j\frac{{\Delta}_{q,j}}{2}\hat{\sigma}_{z, j}  + g_j(\hat{\sigma}_{+, j}\hat{d} + \hat{\sigma}_{-, j}\hat{d}^{\dagger}).
\label{eq:sysHJC}
\end{align}
Here, $\hat{d}$ and $\hat{\boldsymbol\sigma}_j$ are cavity field and qubit Pauli operators {respectively}, and $\epsilon(t)$ describes the amplitude of a coherent drive {applied} at carrier frequency $\omega_d$.  We are in a frame rotating with $\omega_d$: $\Delta_c=\omega_c-\omega_d$ and ${\Delta}_{q,j}= \omega_{q, j}-\omega_d$ are the cavity and qubit detunings respectively.  The cavity qubit coupling, with strength $g_j$, is treated in the rotating wave approximation.

The JC Hamiltonian is accurate for weak readout pulses $\epsilon(t)$, where the role of other qubit energy levels can be ignored \cite{Govia2016, Khezri2016, Malekakhlagh2020}. If the qubit-resonator detuning $\delta_j = \omega_{q, j}-\omega_c$ is large, the cavity population remains below a critical photon number $\expect{\hat{d}^{\dagger}\hat{d}}\ll {\rm min}(|\delta_j/2g_j|^2)$ for a sufficiently weak drive. Eq.~\eqref{eq:sysHJC} can then be perturbatively transformed \cite{Hutchison2009}, {yielding what} we refer to here as the dispersive model:
\begin{align}
\hat{\mathcal{H}}_D = &\Delta_c \hat{d}^{\dagger}\hat{d}  +\epsilon(t)(\hat{d}+\hat{d}^\dagger)
+ \sum_j\frac{\tilde{\Delta}_{q,j}}{2}\hat{\sigma}_{z, j} 
+ \chi_j\hat{\sigma}_{z, j}\hat{d}^{\dagger}\hat{d} \nonumber \\
&+\sum_{jk}J_{jk}\hat{\sigma}_{-,j}\hat{\sigma}_{+,k},
\label{eq:sysHd}  
\end{align}
valid to second order in $g_j/\delta_j$. Here $\chi_j =g_j^2/\delta_{j}$ describes the dispersive shift of the cavity frequency due to the state of qubit $j$ and $\tilde{\Delta}_{q, j} = {\Delta}_{q,j} +\chi_j$ is the renormalized qubit detuning. The effective coupling between qubits via their shared cavity is $J_{jk}={g_jg_k}(\delta_{j}+\delta_{k})/{2\delta_{j}\delta_{k}}$, a manifestation of cross-talk.


We denote the multi-qubit state $\ket{\psi(t)} = \sum_z c_z(t)\ket{z}$, where $\ket{{ z}}= \ket{z_j}^{\otimes N_q}$ represents the $z$-basis state of each qubit as a binary digit 
($\hat{\sigma}_z\ket{0/1} = \mp\ket{0/1})$).  We consider the standard measurement process here, where the cavity is initially in the vacuum state and a coherent drive is applied at $t=0$: $\epsilon(t) = \epsilon_0\Theta(t)$. The $X$-quadrature of the cavity follows a unique trajectory for each multi-qubit state, and by measuring $\expect{\hat{d}+\hat{d}^\dagger}(t)$ one seeks to determine $\ket{\psi(0)}$. In this work, we specialize to the case of two-qubit readout ($z = \{ 00, 01, 10, 11 \}$), which is later seen to be a non-trivial classification task. We consider both the dispersive and JC models with following parameters in units of the cavity decay rate $\kappa$:
$\Delta_c = 0$,
$\epsilon_0 = 2$,
$\chi_1=1.8$, 
$\chi_2=1.3$,
$\Delta_{q,1}=180$,
$\Delta_{q,2}=130$,
$g_j/\delta_j=10^{-1}$, and include additional qubit decay with rate $\gamma_h=10^{-2}$.  These are all physically plausible parameters for current superconducting circuit implementations of this system.

\begin{figure}[t]
\captionsetup[figure]{labelformat=empty}
\includegraphics[width=0.48\textwidth]{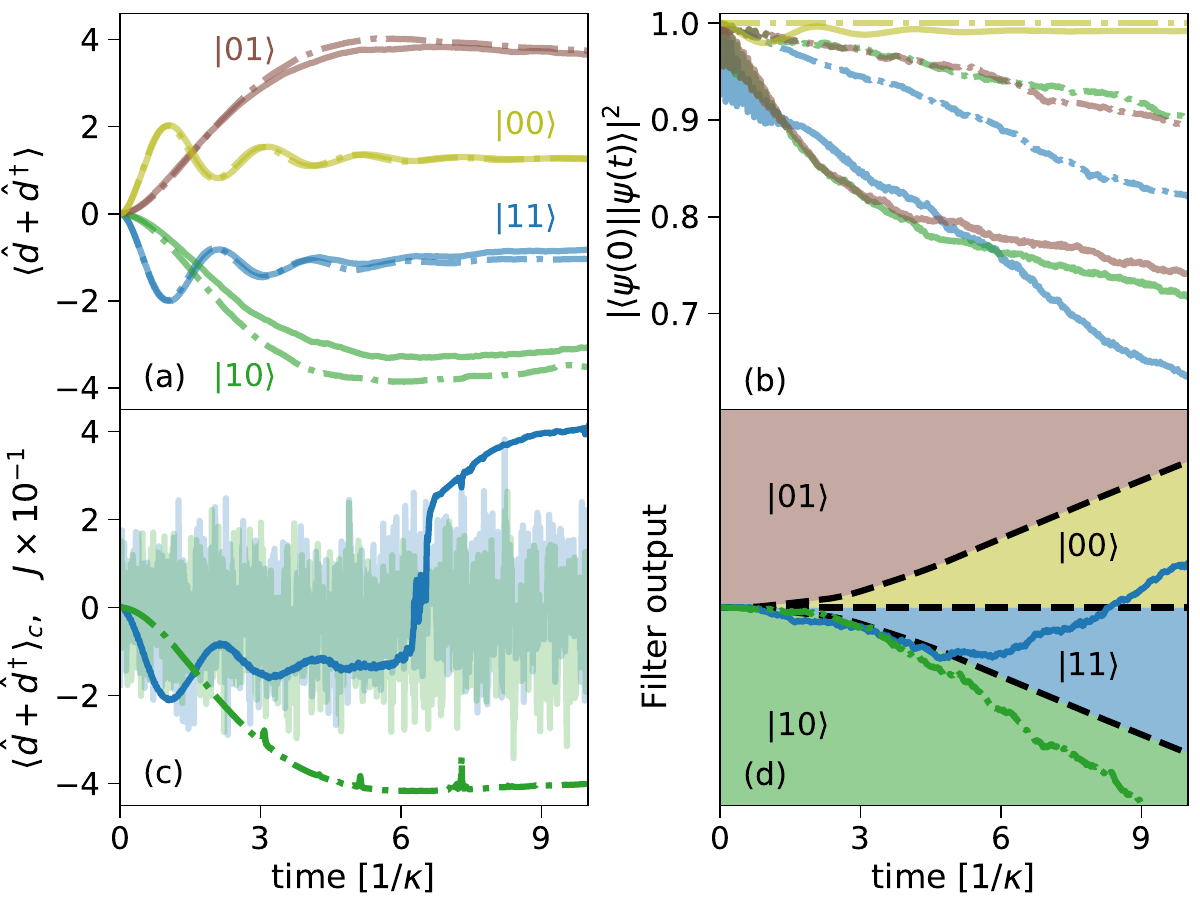}
\caption{(a) Ensemble-averaged cavity field quadrature during readout, for each initial qubit state $\ket{\psi(0)} = \ket{z}$.  (b) Ensemble-averaged decay of the initial qubit state occupation during measurement: $|\langle \psi(0)|\psi(t)\rangle|^2$.  In all plots, results for the dispersive and JC models are shown in dash-dotted and solid lines, respectively, and  $|\psi(0)\rangle$ for each curve is indicated via the colors labeled in (a).  (c) and (d) demonstrate the matched filtering of dispersive measurement signals.  In (c) we show sample readout cavity quadratures for $\hat{\mathcal{H}}_{JC}$ initialized in $\ket{11}$ {(blue)} and $\hat{\mathcal{H}}_{D}$ in $\ket{10}$ {(green)}.  The corresponding  measurement signals $J$ are also plotted, reduced by an order of magnitude for visibility. The subsequent filtered readout signals $y(t)$ are shown in (d), as computed via an ideal $Q\to\infty$ matched filter. The filtered output is classified according to which expected  bin it falls in, which are labelled with their associated quantum state.  All figures are for the parameters in text.}
\label{fig:filterDyn}
\end{figure}

A thorough discussion of the joint dispersive measurement process is contained in Appendix \ref{app:readout}; in the remainder of this section we summarize the salient results. Figure~\ref{fig:filterDyn}(a) depicts the expected readout cavity evolution $\expect{\hat{d}+\hat{d}^\dagger}(t)$ for each initial qubit state $\ket{z}$ in the measurement basis under the dispersive ($\hat{\mathcal{H}}_D$) and JC models ($\hat{\mathcal{H}}_{JC}$).  In both cases, the cavity evolves to distinct steady-states over a timescale set by $\kappa$.  The difference between these models is manifest in the qubit evolution shown in Fig.~\ref{fig:filterDyn}(b).  Here we plot the expected probability that the system will be measured to be in the multi-qubit state it was prepared in: $|c_z(t)|^2 = |\braket{z}{\psi (t)}|^2$, for $\ket{\psi(0)}=\ket{z}$.  The decay of initially excited states can be seen to be significantly faster for the JC model. In the  dispersive model the qubit state evolution is due to $J_{12}$ and $\gamma_h$, {and} the corresponding {timescales} are taken to be slow relative to the system dynamics. As $J_{12}/\kappa,\gamma_h/\kappa\to 0$ the measurement process becomes QND since the qubit state is conserved, where in a perfect QND measurement $|c_{z}(t)|^2 =1$.  In contrast, the JC interaction $\propto g_j$ does not commute with $\hat{\sigma}_{z, j}$, and this fast Hamiltonian evolution causes information about the initial qubit state to be lost more quickly during measurement. 

We consider the situation where the output cavity field  $X$-quadrature is continuously monitored via homodyne detection; the QS ($\hat{\mathcal{H}}_{S} = \{\hat{\mathcal{H}}_{D},\hat{\mathcal{H}}_{JC}\}$) then  evolves under the stochastic master equation (SME) \cite{Jacobs2006, Gambetta2008}:
\begin{align}
\dot{{\hat\rho}} = 
 &-i[\hat{\mathcal{H}}_{S}, \hat{\rho}]  
+\gamma_h\sum_j \mathcal{D}[\hat{\sigma}_{-,j}]\hat{\rho} \nonumber \\
&+\kappa \mathcal{D}[\hat{d}]\hat{\rho} 
+\sqrt{\kappa}\mathcal{M}[\hat{d}]\hat{\rho}~\xi(t)
\label{eq:syssme}
\end{align}
In the above, the  dissipative and measurement superoperators are respectively 
$\mathcal{D}[\hat{O}]\hat{\rho} = \hat{O}\hat{\rho}\hat{O}^\dagger -\frac{1}{2}\{\hat{O}^\dagger\hat{O}, \hat{\rho}\}$,
$\mathcal{M}[\hat{O}]\hat{\rho} = \hat{O}\hat{\rho} + \hat{\rho}\hat{O}^\dagger - \expect{\hat{O} + \hat{O}^{\dagger}}_c\hat{\rho}$. 
The SME describes {the evolution of} the QS $\hat{\rho}$ \textit{conditioned} on the observed measurement outcome $J(t)$.
The outcome of a measurement is the continuous classical current
\be
J(t)=\sqrt{\kappa}\langle \hat{d}+\hat{d}^\dagger\rangle_c({t}) + \xi(t)
\label{eq:Jhom}
\ee
where $\xi(t)$ is white noise, arising from fundamental quantum uncertainty in the cavity state:  ${\expect{\xi(t)}} = 0$, ${\expect{\xi(t)\xi(t'))}} = \delta(t-t')$.  In the above,  the subscript $c$ denotes expectation values taken with respect to the conditional state $\hat{\rho}$. 

These quantities evolve stochastically during individual measurements: samples of $J(t)$ and $\langle \hat{d}+\hat{d}^\dagger\rangle_c({t})$ are depicted in Fig~\ref{fig:filterDyn}(c).  The measurement signals $J(t)$ are dominated by noise $\xi(t)$, and the measurement process produces backaction on the quantum state, resulting in, for example, the sudden jump seen for the JC sample.
By taking the ensemble average of many measurement records however, one recovers the unconditional system dynamics of Fig.~\ref{fig:filterDyn}(a), described in Appendix \ref{app:readout}.  Our quantum measurement data is constructed by numerically integrating the SME of Eq.~\eqref{eq:syssme} from initial states $\hat{\rho}(0) = \ketbra{0, z}{0, z}$ using QuTip \cite{Johansson2013}.  Each trajectory $q$ has a unique noise record $\xi^{(q)}(t)$ and thus conditional expectation values $\expect{\hat{O}}^{(q)}_c$ and measurement signal $J^{(q)}(t)$. 



The individual measurement currents $J^{(q)}(t)$ have a small signal-to-noise ratio (SNR) due largely to the additive white noise term in Eq.~\eqref{eq:Jhom}, obscuring the relevant conditional evolution, particularly when the cavity photon number is kept low to keep the measurement in the QND regime.  In particular, for the chosen parameters the steady-state measurement current SNR are  $-6.8\,\rm{dB}$ and $-7.1\,\rm{dB}$ for the dispersive and JC systems respectively.  Further signal processing is thus needed to extract the underlying initial qubit state;   conventionally, this is done by constructing a matched filter (MF) from a large set of measurement currents for which the initial qubit state is known: $U(t) = \{J^{(q)}(t)\}$ \cite{Gambetta2007, Heinsoo2018}. 
Specifically, a MF is a linear filter with kernel $h(\tau)={\expect{U^*(t-\tau)}}$, the conjugated and time-reversed mean of the training set.  For an input consisting of a signal plus white noise, the output approaches the auto-correlation function of the signal: 
$y(t)=\int^t d\tau~h(t-\tau) u(\tau)=\int^t d\tau{\expect{U^*(\tau)}}u(\tau)$.
This maximizes the SNR; a MF is the optimal linear filter for distinguishing signals, such as we study here, with additive noise.  For the two-qubit readout system, the matched filter is constructed by averaging the absolute value of all four sets of mean outputs, which maximizes the overall fidelity with which initial states can be distinguished. The more sample trajectories used to `train' the MF, the better $h(\tau)={\expect{U^*(t-\tau)}}$ is expected to represent the  underlying signal, improving the filter performance.

The filter is used to define an expected bin for each filtered signal as a function of time, and quantum states are classified according to which bin they fall into.  The MF classification process is depicted in Fig.~\ref{fig:filterDyn}(d); the filter is seen to remove the white noise from the homodyne signals, and the sample from the dispersive system falls into the correct bin reasonably quickly.  The readout process has a more significant influence on the qubit state in the JC system, which in this case causes the first qubit to decay at $t\sim 6/\kappa$.  This results in the cavity state suddenly jumping as well, and the filtered output falls into the wrong bin at later times as a result. This loss of initial state information, at rates depicted in Fig.~\ref{fig:filterDyn}(b), places a limit on maximum fidelity with which signals can be classified, since the task is to learn $\ket{\psi(0)}$, not $\ket{\psi(t)}$.



\section{A Kerr Network Reservoir Computer}
\label{sec:KerrRC}

\subsection{Proposal and Overview}
\label{subsec:proposal}

\begin{figure}[t]
\subfloat{\includegraphics[width=0.48\textwidth]{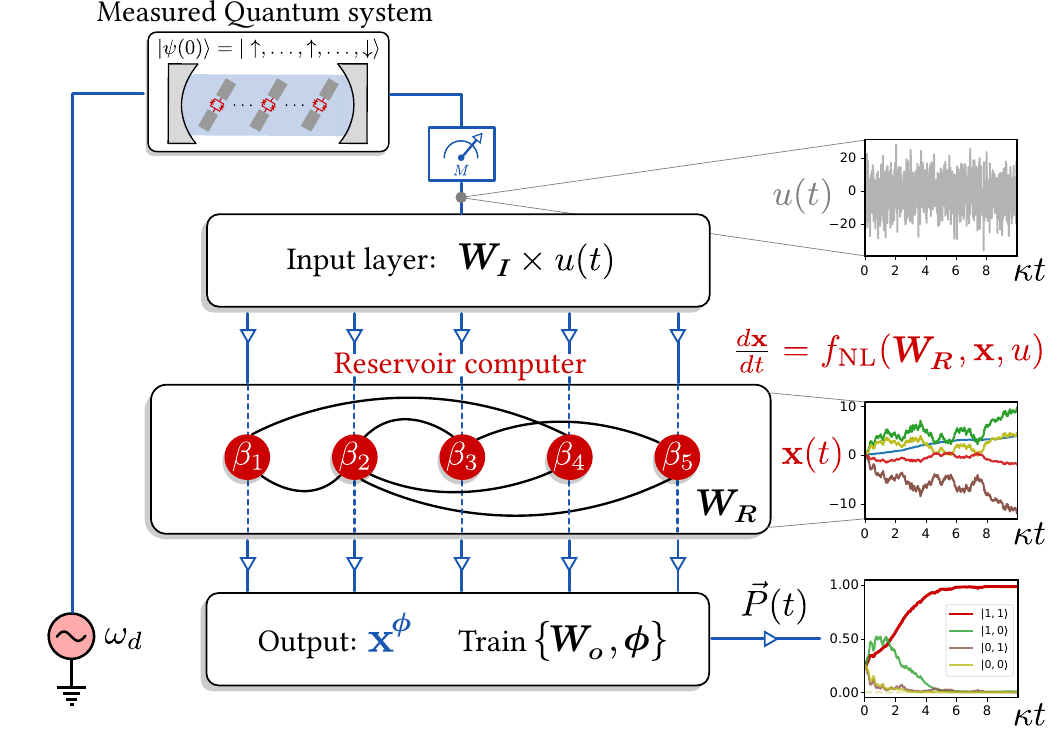}} 
\caption{A schematic of our proposal to use a hardware RC to process quantum measurement signals, here shown for $K=5$ Kerr nodes.  The QS is interrogated, and the resulting measurement record $u(t)$ is input to the reservoir.  The input layer $\bm{W}_I$ randomly couples this signal to each node of the Kerr network, whose subsequent dynamics are a nonlinear function of $u(t)$ and the network itself via Eq.~\protect\eqref{eq:RCeom}.  The output layer performs classification by measuring learned linear combinations of nodes to compute the probability that the input history corresponds to each underlying quantum state.}
\label{fig:scheme}
\end{figure}

Having discussed the conventional approach to quantum state measurement, we will now describe our proposed RC approach, depicted schematically in Fig.~\ref{fig:scheme}. Generally, the RCs task is to classify the state $|\psi (0)\rangle$ of the target QS, based on a quantum measurement that results in a noisy readout signal containing information about this initial state. We specialize to the QS described in the previous section and thus homodyne measurement current of Eq.~\eqref{eq:Jhom} in this article, but note that our approach can be applied to entirely different systems and measurement modalities.

Instead of conventional processing via room-temperature electronics and a software backend, the readout signal is fed into a hardware reservoir processor via a fixed linear input matrix $\bm{W}_I$. In our specific realization, this RC consists of a network of Kerr nonlinear oscillators. These nodes then evolve according to this input and internal structure defined by a connectivity matrix $\bm{W}_R$ and nonlinearity vector $\bm{\Lambda}$.  This results in a complex dynamical mapping:
the physical state of the RC $\bm{x} (t)$ is a high-dimensional nonlinear function of the input history $\bm{u}(\tau<t)$. The output of the computation is a linear combination of the RC nodes $\bm{y}(t) = \bm{W}_o \, \bm{x}(t)$, where the matrix $\bm{W}_o$ can be trained such that $\bm{y}(t)$ approximates a desired function of the input $\bm{F}(\bm{u} (\tau<t))$. 

Only this output layer is trained; the training step is thus a simple convex optimization problem with a limited number of parameters, which is guaranteed to converge \cite{Jaeger2009, Larger2017}. This is in contrast to the `vanishing gradient problem' that plagues training of other neural networks~\cite{Goodfellow2016}. Although it may seem like training only output weights $\bm{W}_{o}$ would lead to inferior results for time-series processing, head-to-head comparisons between state-of-the-art recurrent neural networks and RCs show surprisingly similar performance \cite{Vlachas2018,Bompas2020, Bollt2021}, despite RC training protocols being $10^3$ to $10^6$ times quicker. Since the internal structure is not optimized, RCs can be quickly retrained for different tasks, resulting in them being a powerful and generalizeable computational tool \cite{Appeltant2011, Griffith2019, Tanaka2019, Canaday2020}. 

For our task, the RC is trained by preparing the QS in known initial states and then performing readout.  This allows one to optimize linear combinations of Kerr RC node quadratures, which are measured such that the output of the RC correctly assigns the quantum state. After training, the classifier continuously outputs the probability $\bm{P}(t)$ that the measurement record up to the the present time corresponds to each underlying initial quantum state. The RC thus replaces the filtering and binning step of the quantum measurement process; its efficacy in doing so will be discussed in Sec.~\ref{sec:RCproc}.

The Kerr network reservoir we propose may be implemented in a superconducting circuit platform via a network of coupled Josephson Parametric Oscillators (JPOs). These JPOs can be either single Josephson junctions or composite elements such as  Superconducting Nonlinear Asymmetric Inductive eLements (SNAILs) \cite{Frattini2017}. The recurrent connections can be flexibly generated by coupling the JPOs to a common electromagnetic resonator mode. Variants of such networks have been considered as hardware for superconducting quantum annealers \cite{Puri2017,Nigg2017,Onodera2020} and for stabilization of multi-qubit entanglement \cite{Aron2016,Kimchi-Schwartz2016}. Such an RC then could be integrated with the QS to be measured, sharing a cryogenic environment. For the present work we require that the JPOs are in the weakly nonlinear regime, that their scale of nonlinearity is much smaller than their dissipation. This is the regime of Josephson parametric amplifiers \cite{Roy2016}.  Another particularly interesting platform to realize this hardware RC is an optical Kerr network, which would then be well-suited to the readout of optical QSs.  An important advantage of either of these implementations of Fig.~\ref{fig:scheme} is that the RCs will operate at the timescales of the measured quantum system, faster than conventional FPGA-based electronics and potentially allowing for real-time analog processing.

\subsection{Kerr Network Reservoir Computer Model}
\label{subsec:KerrRCmodel}

The reservoir, consisting of a network of coupled Kerr-nonlinear oscillators, is described by the master equation:
\be
\dot{\hat{\rho}}_{RC} = -i[\hat{\mathcal{H}}_{RC}, \hat{\rho}_{RC}] +\sum_k \gamma_k \mathcal{D}[\hat{b}_k]\hat{\rho}_{RC}
\ee
where the governing Hamiltonian $\hat{\mathcal{H}}_{RC}$ takes the form:
\begin{align}
 \hat{\mathcal{H}}_{RC} =& \sum_k \Delta_k \hat{b}^\dagger_k \hat{b}_k - \frac{\lambda_k}{2} \hat{b}^{\dagger 2}_k \hat{b}^2_k
     + \sum_{kl}  \varg_{kl}  \hat{b}^\dagger_k \hat{b}_l \nonumber \\
    & + \sum_{km}  i\varepsilon_{km}  (u_m(t)\hat{b}^\dagger_k - u^\dagger_m(t)\hat{b}_k)   
\end{align}
The input to the RC is a collection of signals ${\bm u(t)}$ with a common carrier frequency $\omega_{d}$.  Each nonlinear oscillator is described by a field operator $\hat{b}_k$, with detuning $\Delta_k$ from the carrier frequency and Kerr nonlinearity $\lambda_k$.  $g_{kl}$ and $\varepsilon_{km}$ are the linear couplings between oscillators and to the input respectively, and  $\gamma_k$ is the energy decay rate. Inter-oscillator couplings can be generated by cavity-mediated interactions~\cite{Aron2016, Kimchi-Schwartz2016} or through parametric means~\cite{Onodera2020}.
 
The evolution of the field amplitude from each node is given by the Heisenberg equation of motion:
\begin{align}
 { \expect{\dot{\hat{b} }_k }} = &-(i\Delta_k +\frac{\gamma_k}{2}) \expect{\hat{b}_k} +i \lambda_k \expect{\hat{b}^{\dagger}_k \hat{b}^2_k} \nonumber \\
&-i \sum_l g_{kl}\expect{\hat{b}_l} +\sum_m\varepsilon_{km}u_m(t)
\label{eq:RCeomq}   
\end{align}

We will consider Kerr networks where the field amplitudes are sufficiently large that they are in the classical regime. 
{For $c\in\mathbb{R}^+$, defining scaled drive strengths $\tilde{\varepsilon}_{km}=\sqrt{c}{\varepsilon}_{km}$, nonlinearity $\tilde{\lambda}_k = \lambda_k/c$, and introducing $\beta_k \equiv \sqrt{c}\expect{\hat{b}_k}$, it can be shown that $\expect{\hat{b}_k^{\dagger}\hat{b}_k^2} = |\beta_k|^2\beta_k + O(\frac{1}{c})$. Heuristically, this indicates that for $c\to\infty$, where the nonlinearity grows weaker and the `classical' occupation $|\beta_k|^2 = c|\expect{\hat{b}_k}|^2$ becomes simultaneously larger,
quantum correlations captured in higher-order moments can be neglected.
In this case, Eq.~\eqref{eq:RCeomq} becomes:
\be
2\dot{\beta}_k/\gamma = -\beta_k +i \Lambda_k |\beta_k|^2 \beta_j
-i \left(\bm{W}_{R}\cdot\bm{\beta}\right)_k  + \left(\bm{W}_{I}\cdot \bm{u}\right)_k
\label{eq:RCeom}
\ee
Without loss of generality, we have chosen the nodal decay rates to be identical $\gamma_k = \gamma$ to define the dimensionless parameters, familiar to the RC framework:
\begin{align}
    W_{R, kk} = 2\Delta_k/\gamma, \,\,
    W_{R, kl} = 2g_{kl}/\gamma, \nonumber \\
    W_{I, kl} = 2\tilde{\varepsilon}_{kl} /\gamma, \,\,
    \Lambda_k= 2\tilde{\lambda}_k/\gamma.
\end{align}
}
Eqs.~(\ref{eq:RCeom}) defines the set of ODEs governing the RC response to a given input signal ${\bm u}(t)$ for a $K$-node RC.

As mentioned in Sec.~\ref{subsec:proposal}, the philosophical underpinning of reservoir computing is that if one has a sufficiently complex and high-dimensional system, there is no need to optimize its many internal parameters for a specific computational task.  As such, here we consider random Kerr networks,  whose internal structure and dynamics, specified by $\bm{W}_I$, $\bm{W}_R$, $\boldsymbol\Lambda$ and $\boldsymbol\gamma$ via Eq.~\eqref{eq:RCeom}, are set randomly and \textit{not individually optimized}.  Instead, the RC properties are controlled by the scale-independent hyperparameters $\{\gamma, \alpha, \bar{\Lambda}, \mu\}$, where:
\begin{align}
     &W_{I,kl} \in [-\mu, \mu], \,\, 
    \Lambda_k \in [0, 2\bar{\Lambda}], \nonumber  \\
    &W_{R,kl} \propto [-1,1]~{\rm s.t.}~\alpha = \lambda_{max}(\bm{W}_R),
    \label{eq:netparams}
\end{align}
and $\lambda_{max} (\bm{W}_R)$ refers to the maximum singular value of the connectivity matrix $\bm{W}_R$. Here $[a,b]$ defines a uniform distribution with probability density within the limits $(a,b)$, so that the various Kerr RC internal parameters are obtained by randomly sampling appropriate uniform distributions. We constrain the ranges of these hyperparameters to be compatible with the proposed hardware realizations, while also importantly enabling desired RC evolution properties of fading memory, separability and nonlinearity; their selection is discussed in more detail in Sec.~\ref{subsec:RCdesign}. We will see throughout this work that the generic behaviour of an RC is well-quantified by its hyperparameters, and that performance is robust to both network structure and variations in these values. We have also introduced significant variation into node decay rates $\gamma_k$ and observed that the RC performance is again unchanged.

The output of the Kerr RC is a linear combination of the measured quadrature variables $x_k^{\phi_k}(t)$ of its nodes. 
These quadrature variables are defined in terms of the $2K$-element vector of complex RC node amplitudes:
$
\bm{x} = \sqrt{2} \left( {\rm Re }\{ \beta_1 \}, {\rm Im }\{ \beta_1 \},
\ldots,{\rm Re} \{\beta_K\},{\rm Im} \{\beta_K\}) \right)^T
$
and the $K\times 2K$ measurement matrix
\be
\bm{C}(\bm{\phi}) = 
\begin{pmatrix}
\cos \phi_1 & \sin \phi_1 & \ldots & 0  & 0 \\
\vdots & \vdots & \ddots & \vdots & \vdots \\
0 &  0 & \ldots & \cos \phi_K & \sin \phi_K 
\end{pmatrix}
\ee
such that
$
\bm{x}^{\bm{\phi}} = \bm{C}(\bm{\phi}) \bm{x} 
$ 
defines the RC node quadratures measured. The orthogonal quadrature then constitutes the `hidden variables' which do not contribute directly to the output.  The RC output $y(t)$ can then be expressed
\be
\bm{y}(t)=\bm{W}_o\bm{x}^{\bm{\phi}}(t)=
\bm{W}_o\bm{C}(\bm{\phi}) \bm{x}(t)
\label{eq:RCout}
\ee
where $\bm{W}_o$ is a $C \times K$ dimensional matrix of output weights, which together with the $K$ element vector $\bm{\phi}=(\phi_1\ldots \phi_K)$ of measurement angles defines  a total of $(C+1)\cdot K$ parameters to optimize in the RC training.

Finally, specific to the task of classifying $C$ states of the QS, it is appealing to map the RC output to a probability that the input to the RC corresponds to the QS initialized in $\ket{\psi(0)}=\ket{z}$. This is achieved by applying the `softmax' function (Boltzmann distribution) to the trained RC output:
\be
P_j(t) = e^{{ y}_j(t)}/\sum_k e^{{ y}_k(t)},
\label{eq:Pout}
\ee
which maps the RC outputs to mutually exclusive probabilities that sum to unity.

Details of and background on RC training are discussed in Appendix \ref{app:Train}; we briefly summarize some of the salient aspects here. We first construct a training data set consisting of $Q$ measurement currents  $U(t) = \{J^{(q)}(t)\}$ for a measurement period $\tau_m$ for associated known initial states $\bm{ y}^{\star (q)}= z^{(q)}$ ($q=1,\cdots, Q$). $M$ measurements are conducted for each initial state by integrating the SME of Eqns.~\eqref{eq:syssme} and \eqref{eq:Jhom}.  The RC node trajectories are simulated by solving Eq.~\eqref{eq:RCeom}, and the resultant states are used to minimize the multinomial cross entropy cost function $\mathcal{L}_x$ (Appendix \ref{app:Train}), optimizing $\bm{ W}_o$ and $\bm{\phi}$ for the training set.  We consider small RCs ($K=2-5$) and small training sets ($Q < 100$). Therefore, the optimization of the cross entropy loss function on a digital computer is a quick convex optimization problem with a small set of output parameters. We stress that this network size is orders of magnitude smaller than typical RCs, a choice made for hardware-realizability. After training, the continuous RC output from Eqs.~\eqref{eq:RCout} and \eqref{eq:Pout} provides the probability $P_z(t)$ that the observed record $J^{(q)}(t)$ up to the current time corresponds to the initial state $z^{(q)}$.

\section{Reservoir processing of quantum measurement}
\label{sec:RCproc}

We now analyze the performance of the proposed Kerr network RC on the two-qubit readout task described above. We focus on the ability of an RC trained with a small labelled training set to classify a much larger set of unknown test signals (quantum states) from either the dispersive or JC system ($\hat{\mathcal{H}}_S=\{\hat{\mathcal{H}}_D,\hat{\mathcal{H}}_{JC}\}$ in Eq.~\eqref{eq:syssme} respectively). {An obvious metric to evaluate the performance is} the `classification accuracy' $C_{\mathcal{F}}(t)$, {which} refers to the fraction of test signals the RC correctly classifies at a given readout time $\tau_m$. However, classification accuracy does not increase monotonically with readout time; as indicated in Fig.~\ref{fig:filterDyn}, the initial qubit state stochastically evolves and can be lost during the measurement process, {an outcome} that is particularly likely for the JC system.  After this point $J(t)$ no longer faithfully provides information about $\ket{\psi(0)}$, imposing a steadily decreasing ceiling on $C_{\mathcal{F}}(t)$ (as can be seen in Fig.~\ref{fig:claFD}).

In practice, one would simply stop the measurement process and RC computation at the time $C_{\mathcal{F}}(t)$ peaks; this time is consistent for a given QS being measured.  Thus, we define the `Classification Fidelity' $\mathcal{F}$ of an RC or filter to be this peak accuracy
\be
\mathcal{F} = {\rm max}_t \{C_{\mathcal{F}}(t)\}
\ee
and will seek to optimize this metric when considering RC design in Sec.~\ref{subsec:RCdesign}.  Since the maximum possible $C_{\mathcal{F}}(t)$ decays with time, $\mathcal{F}$ combines both speed and accuracy of classification.

We compare the RC performance against that of conventional filtering. In particular, we consider a boxcar filter (BF), which amounts to integrating $J(t)$ to remove noise, a matched filter (MF) constructed using a training set of size $Q$ of labelled homodyne currents, and a MF constructed using the analytic solution $\expect{\hat{d}+\hat{d}^\dagger}_z$ to $\hat{\mathcal{H}}_D$.  For the dispersive system, this corresponds to the $Q\to\infty$ limit of the MF constructed from homodyne currents.  
{In many realistic scenarios, the system model and parameters are not known exactly or are subject to experimental drift, and attempting to construct an analytic MF is impractical.  Instead, either the BF or a finite $Q$ MF is used because they (like the RC) do not require a model of the underlying system; for the MF this requires regularly producing large training sets every time the device is re-calibrated.} 

\subsection{Reservoir Dynamics and Classification Fidelity}
\label{subsec:RCdyn}

\begin{figure}[h]
\captionsetup[figure]{labelformat=empty}
\includegraphics[width=0.47\textwidth]{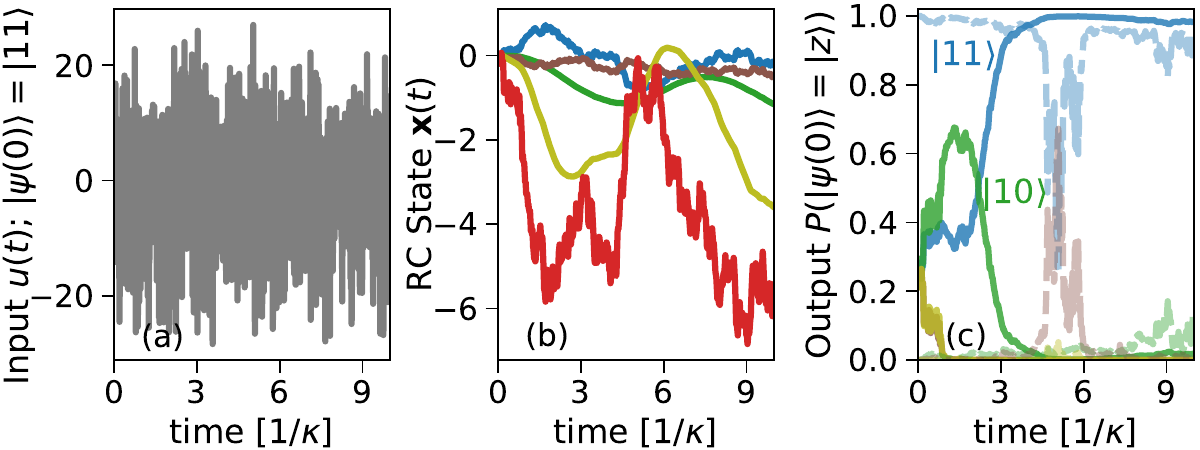}
\caption{A demonstration of the RC classification process. (a) depicts a sample quantum signal $u(t)=J^{(q)}(t)$ generated by the joint dispersive readout under $\hat{\mathcal{H}}_D$ of the initial state $\ket{\psi(0)}=\ket{11}$.  This drives a $K=5$ node RC, producing the node state dynamics $x^{\phi_k}_k(t)$ shown in (b). The classification result is in (c): the RC, previously trained with $Q=40$ readout signals, continuously updates the probability that the initial state was a specific state in the measurement basis.  These probabilities are shown in solid lines with the corresponding color-scheme from Fig.~\protect\ref{fig:filterDyn}.  The decomposition of the true quantum state conditional on the measurement record up to time $t$ is depicted in lighter dashed lines; the RC successfully classifies the initial state after $\sim 3/\kappa$. Note that $|c_{01/00}|^2(t)$ (and associated $P$)  are vanishingly small
.}
\label{fig:claDyn}
\end{figure}

\begin{figure}[h]
\captionsetup[figure]{labelformat=empty}
\includegraphics[width=0.47\textwidth]{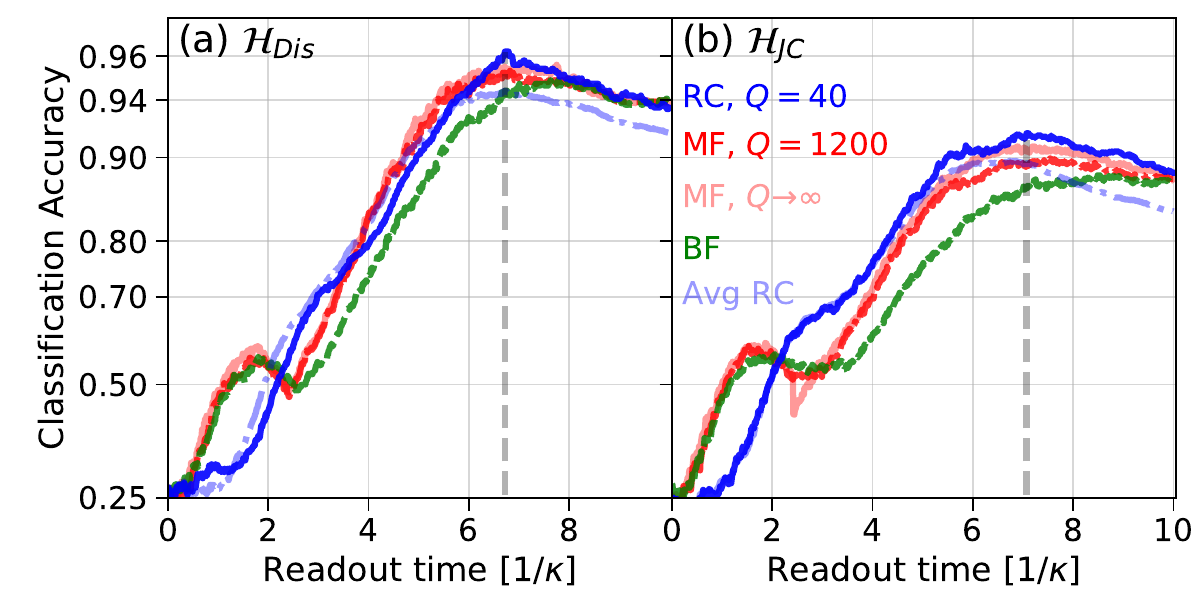}
\caption{Classification accuracy vs readout time on a $1200$ signal test set for the dispersive system in (a) and the JC system in (b).  In both cases, the RC of Fig.~\protect\ref{fig:claDyn} (solid blue) is trained with $Q=40$ measurements and the optimal readout time is indicated.  The average classification fidelity of 10 different random RCs sharing the same hyperparameters is also shown in light dash-dotted blue.  For comparison purposes we also show the classification accuracy of a BF (dashed green), analytic MF (light red), and  $Q=1200$ MF (dash-dotted red).
}
\label{fig:claFD}
\end{figure}

The response of a representative $K=5$ node RC to a quantum readout signal from the dispersive system is depicted in Fig.~\ref{fig:claDyn}.  Here (and in Fig.~\ref{fig:claFD} to follow) the RC has hyperparameters $\gamma = 0.7\kappa$,
$\alpha = 1.9$,
$\bar{\Lambda} = 5\times 10^{-2}$,
and 
$\mu = 5$.  
The RC is driven with measurement signals $u(t) = J^{(q)}(t)$ with a total measurement time  $\tau_m = 10/\kappa$ and has previously been trained with $Q=40$ readout signals (recall that this is $M=10$ samples for the four states) to optimize $\{\bm{W}_o, \bm{\phi} \}$.  In Fig.~\ref{fig:claDyn}(a) we show one such noisy signal for an unknown (to the RC) quantum state $\ket{\psi(0)}=\ket{11}$.  Once the readout drive is turned on, the noisy input $u(t)$ acquires a non-zero mean.  This drives the network away from its rest state $\bm{x}=0$ to a non-trivial state in its phase space, as can be seen via the $\bm{x}^{\bm{\phi}}$ trajectories in \ref{fig:claDyn}(b).  
Real time classification is performed by reading out each node continuously: the output ${P}^{(q)}_z(t)$ are shown in \ref{fig:claDyn}(c).  After only $t=3/\kappa$, a timescale shorter than that over which the cavity reaches steady state (see Fig.~\ref{fig:filterDyn}(a)), ${P}^{(q)}_{11}>0.5$, and the RC thus has quickly and correctly classified this sample quantum state. The low measurement SNR while the cavity is being populated is responsible for the RC initially not distinguishing $\ket{10}$ and $\ket{11}$, with ${P}^{(q)}_{10}\sim{P}^{(q)}_{11}$.  The true quantum state 
$|c^{(q)}_z(t)|^2= {\rm tr} \{\hat{\rho} (t)\ketbra{z}{z} \}$
is also depicted, and we see that the RC classification ${P}^{(q)}_{11}(t)$ is robust to the significant probability amplitude fluctuations: the system almost jumps to the state $\ket{01}$ for this specific trajectory, but ${P}^{(q)}_{11}$ remains saturated.

The ability of this RC to classify unknown quantum states is quantified in Fig.~\ref{fig:claFD}, where we evaluate its performance on a test set of $1200$ unknown quantum signals generated from the dispersive and JC SMEs, and compare with that of various linear filters.  The RC is trained as above using a $Q=40$ measurement set for the corresponding QS, while the MF is constructed from a much larger $Q=1200$ set.
We plot the classification accuracy (fraction of test signals classified correctly) as a function of readout time, for both the RC and the filters; it is apparent that the RC is able to rapidly and reliably extract the initial quantum state and thus perform the readout for both QSs. 

For the dispersive system both the RC and the MFs have a classification fidelity $\mathcal{F}>0.96$, achieved for a readout time $\tau_m\sim 6.7/\kappa$.  This fidelity is limited by non-QND dynamics that result in the initial state information being lost.  As previously noted, measuring beyond a model-dependent optimal time thus increases the odds of an incorrect classification, causing the decrease in accuracy seen at longer times for all methods in Fig.~\ref{fig:claFD}.  
Conversely, at shorter readout times the SNR is much lower:  the measurement cavity is still being populated, so the output signal is dominated by {shot} noise. A dip in the MF  performance is seen around $2.5/\kappa$; the expected filtered signals for $\ket{1(0)0}$ and $\ket{1(0)1}$ cross for this specific measurement time, meaning the MF has no information about the second qubit.  This occurs in the BF as well at a slightly later time.  The RC avoids such a problem by mapping the qubit state into its higher-dimensional phase space.  

As described in Sec.~\ref{sec:quantumsystem}, initial state information is lost more quickly for the JC system due to non-QND Hamiltonian evolution.  Despite this, the RC is able to accurately process these quantum measurements, with a classification fidelity of $0.92$, exceeding that of any linear filter.  The optimal measurement time is slightly later, at $\sim 7/\kappa$.  This is due to the decreased SNR for the JC readout task: 
$\sqrt{\kappa}{\expect{ \hat{d}+\hat{d}^\dagger}} $ 
is reduced relative to the dispersive case and there is increased measurement backaction noise (see Fig.~\ref{fig:filterDyn}(a) and (c)).
{
As noted earlier, these measurement currents have SNR of  $-6.8\,\rm{dB}$ and $-7.1\,\rm{dB}$ for the dispersive and JC systems respectively when the cavity is in steady-state.  We can decrease the SNR further by introducing a measurement efficiency $\eta\leq 1$ ($\sqrt{\kappa}\to\sqrt{\kappa\eta}$ in Eqns.~\eqref{eq:syssme} and \eqref{eq:Jhom}).  For both systems the fidelity of the RC classification decreases steadily with increasing relative noise strength, but remains comparable with that of the MF.
}

This performance is in no way unique to the specific random realization of the RC network; the average classification accuracy of 10 random $K=5$ RCs is also shown for both QSs in Fig.~\ref{fig:claFD}.  These RCs have different structure ($\bm{W}_I$, $\bm{W}_R$, $\boldsymbol\Lambda$), but the same hyperparameters and are trained on the same $Q=40$ training set.  The peak classification accuracy occurs at different times for different RCs, resulting in the average curve being artificially flattened.  The average classification fidelity of the 10 different RCs is $0.951$ for the dispersive system and $0.905$ for the JC system (for $Q=100$, this increases to $0.915$), indicating the robustness of the RC approach.  A few of the random networks experience a sharper drop in classification accuracy after the optimal readout time: this is because the RC states corresponding to different input signals are less separated in phase space.  In the next section we explore the role of the RC evolution in its phase space in more detail.


\subsection{Rapid Training}
\label{subsec:rapidcal}

\begin{figure}[h!]
\captionsetup[figure]{labelformat=empty}
\includegraphics[width=0.48\textwidth]{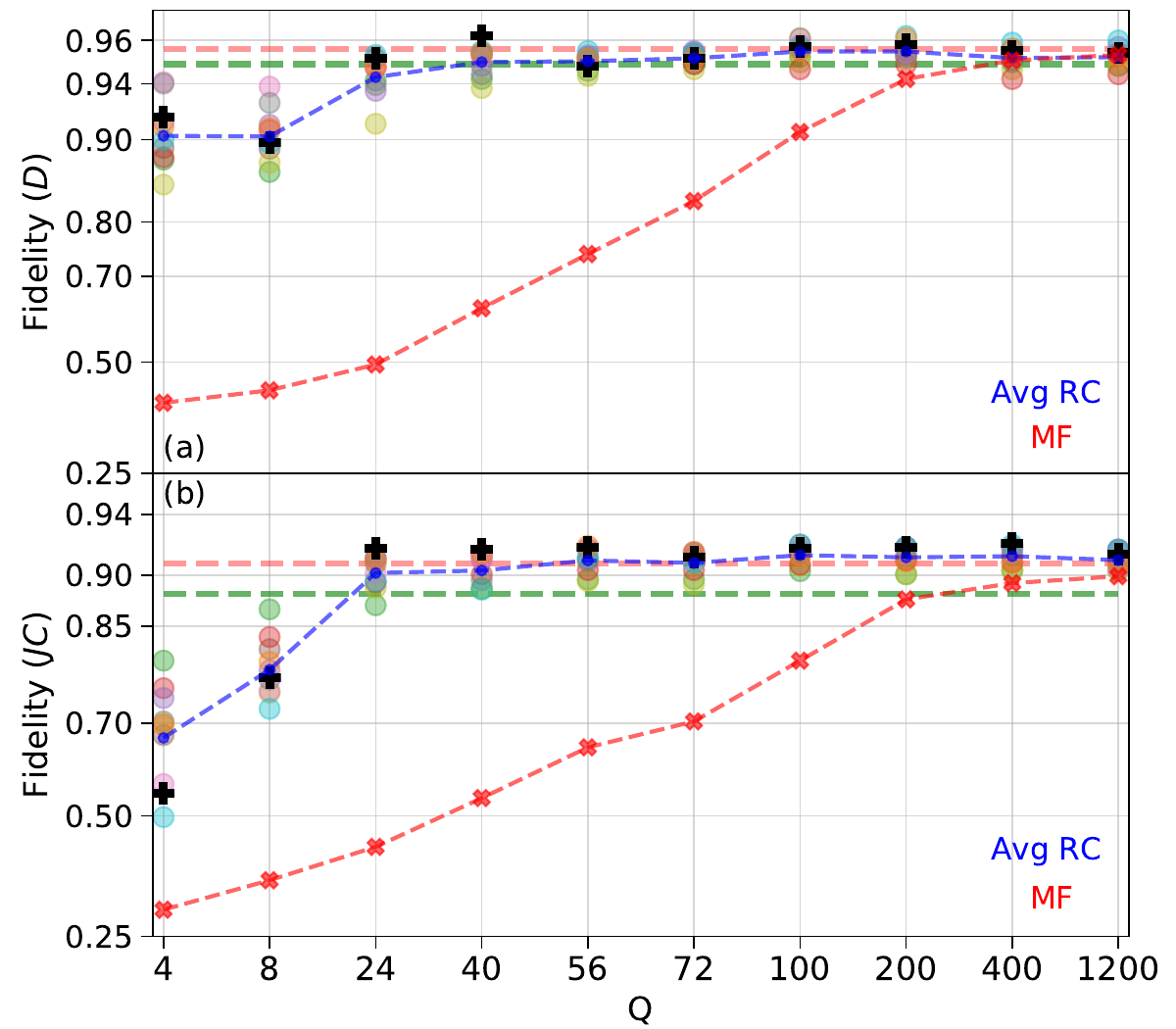}
\caption{Classification fidelity as a function of number of training trajectories for the MF and $K=5$ RCs described in the text.  The dispersive and JC readout systems are in (a) and (b) respectively.  Individual RCs are indicated with different colored circles, and the RC of Fig.~\protect\ref{fig:claDyn} is emphasized with a black plus.  The average RC fidelity is indicated with a dashed blue line, and can be compared with the MF performance, denoted with red crosses and a dashed line.  The performance of the boxcar and analytic MFs do not depend on $Q$, and are denoted with  horizontal dashed green and {pale} red lines respectively.
}
\label{fig:claQ}
\end{figure}

The results of Fig.~\ref{fig:claFD} suggest that an RC is capable of matching the performance of an MF constructed with a much larger training set. We find that RCs hold this training-cost advantage over a MF generally, and demonstrate its dramatic nature in Fig.~\ref{fig:claQ}. Here, we plot classification fidelity as a function of the number of training signals $Q$ 
used to train 10 random $K=5$ RCs from Sec.~\ref{subsec:RCdyn}. Each specific network is indicated in a different color, with the network emphasized in Figs.~\ref{fig:claDyn} and \ref{fig:claFD} highlighted with a plus. The average classification fidelity across the $10$ networks is also shown.  The RCs perform remarkably well at low $Q$, with $Q=4$ readout signals ({just} one for each qubit state) being sufficient to achieve an average fidelity of $0.90$.  This average fidelity increases and the variance in performance across networks decreases as the training set grows, up to $\sim40$ for the dispersive system ($Q\sim 80$ for the JC). Beyond this point, the small fluctuations in fidelity are due to the finite {\em test} set ($1200$ signals) used to calculate $\mathcal{F}$.  The RC thus achieves optimal performance with a small $Q=40-80$ training set for both the dispersive and JC quantum systems, demonstrating its efficacy for rapid measurement calibration.  In contrast, the MF performs very poorly in this regime, needing $Q$ of $O(10^3)$ (note the horizontal axis scale) to converge to the average RC network classification fidelity. 

This significant advantage is particularly relevant for readout of online quantum processors for which  calibrations need to be done regularly{: an} RC-based measurement scheme would thus require far fewer initialization and measurement runs to train than are needed to construct an MF of comparable performance. This can enable a resource-efficient readout calibration system that can also be robustly automatized. The RC network can easily and quickly be re-trained as conditions or even the target QS change, facilitated by the computationally inexpensive software component of training, particularly for small training sets.

{We note that while high-level metrics such as classification fidelity, used widely in the RC literature, are ideal for quantifying the performance of the RC, they do not elucidate the fundamental source of its classification power and fast learning ability. To address this, we carry out a detailed analysis of the phase space dynamics of the Kerr RC nodes in the following section, directly connecting these results to the observed performance.}

\subsection{Phase Space Dynamics}
\label{subsec:phasedyn}

Generally, the effectiveness of the RC approach is attributed to the expressive power of its high-dimensional state space \cite{Pathak2018, Bollt2021}. In the present discussion, the RC under scrutiny transforms a scalar input signal 
into the $2K$-dimensional state space of the RC $\{ \beta_k, \beta_k^* \}$. It is also well-understood that the nonlinearity of the RC plays a crucial role \cite{Appeltant2011,Dambre2012, Griffith2019}, but the underlying mechanism behind how these and other RC properties impact the measurement task are not immediately clear. To perform a fundamental analysis of Kerr RC dynamics and gain unique insight into its previously-described performance, we introduce the Measured Section (MS) of the reservoir. Generally, it is difficult to visualize the dynamics in the high-dimensional RC state space.  Recalling however that we only measure a certain quadrature of the RC oscillators $\bm{x}^{\bm{\phi}}$ (Eq.~\eqref{eq:RCout}), these non-hidden RC nodes evolve in the MS, a $K$-dimensional subspace of the full $2K$-dimensional phase space.  As the $K$ angles $\phi_k$  have been optimized (Sec.~\ref{sec:KerrRC}), this phase-space projection contains the relevant information for the computational task.
 
\begin{figure}[t]
\captionsetup[figure]{labelformat=empty}
\includegraphics[width=0.48\textwidth]{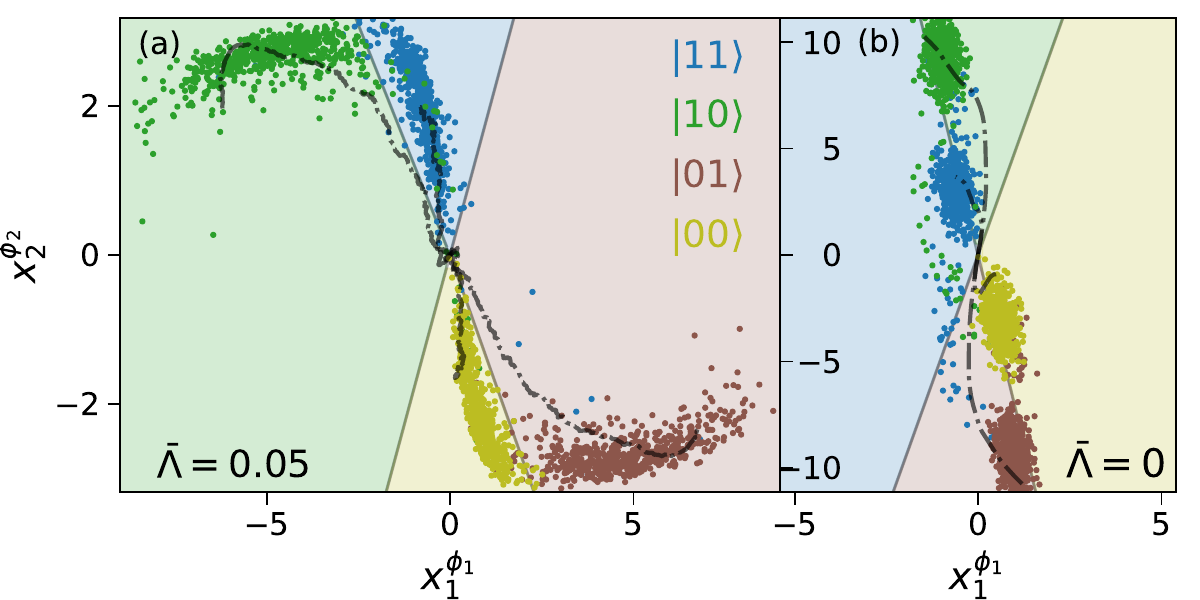}
\caption{(a) Classification dynamics in the measured subspace for a $K=2$ RC with parameters 
$\gamma = 0.2$,
$\alpha = 1.9$,
$\bar{\Lambda} = 5\times 10^{-2}$,
and 
$\mu = 5$. In (b) we show the same network but with ${\boldsymbol\Lambda}=0$. Regions in RC phase space are colored according to their learned classification output, e.g.~the region {in} $(x^{\phi_1}_1, x^{\phi_2}_2)$ space for which $P_{{11}}$ is largest is filled in blue.  The separating lines constructed from the trained $\bm{W}_o$ are also shown.  Dots indicate the final RC state generated by each signal in the test set, which are colored according to their ground truth $\ket{\psi(0)}=\ket{z}$. Therefore, when the colors of the dots and the underlying region are the same, the classification result is correct.
}
\label{fig:phasedyn}
\end{figure}

With the aid of RC dynamics in the MS, we will show that two non-hidden degrees of freedom are sufficient to perform a four-state classification task, \textit{provided} the RC is sufficiently nonlinear. For this low-dimensional ($K=2$) reservoir, the classification process can be conveniently visualized in the MS, as presented in Fig.~\ref{fig:phasedyn}. We consider the response of two RCs, both trained with $Q=40$ measurements, plotting the final reservoir state (at $\tau_m=6.7/\kappa$) in the MS for each signal in a test set of 1200 signals, color-coded with its true $\ket{\psi (0) }$. The RC on the left is nonlinear with $\bar{\Lambda} = 0.05$ and exhibits a $\mathcal{F}=0.96$ on the shown test set, while the RC on the right is linear ($\bar{\Lambda} = 0$) and only attains $\mathcal{F} = 0.75$.



{The training task of learning $\{\bm{W}_o, \bm{\phi}\}$ by minimizing a cost function (detailed in Appendix \ref{app:Train}) is equivalent to finding hyperplanes in the full RC phase space which separate the regions in Fig.~\ref{fig:phasedyn} in a manner that maximally distinguishes the RC states corresponding to different inputs. After optimizing the MS by choosing $\bm{\phi}$, we see that equating rows $j$ and $k$ of the RC output defines the hyperplane in MS separating classes $j$ and $k$: 
$
\left(\bm{W}_o \bm{x}^{\bm{\phi}}\right)_j = \left(\bm{W}_o \bm{ x}^{\bm{\phi}}\right)_k
$. In the $K=2$ case, these hyperplanes are simply lines,
which separate the regions of the MS into four classes, shown in Fig.~\ref{fig:phasedyn} as color-coded sections. Note that due to the symmetry of the dispersive readout classification problem, whereby 
$
{\expect{\hat{d} +\hat{d}^\dagger }}_{11/10} \simeq -{\expect{\hat{d} +\hat{d}^\dagger }}_{00/01}
$,
two pairs of the four separating lines (thin lines in Fig.~\ref{fig:phasedyn}) fall almost on top of each other. 
}

We note that the dynamics of the Kerr RC demonstrate the four properties required of a reservoir,
such that it forms an effective RC~\cite{Canaday2018, Jaeger2009, Dambre2012, Gonon2021}: separation, approximation, fading memory, and nonlinearity. \textit{Separation} is the requirement that different input classes map to distinct regions of phase space, while \textit{approximation} ensures that input series which are close generate RC states which are similarly close.  Both these properties are manifest in Fig.~\ref{fig:phasedyn}, where the final reservoir states for each class fall into separated regions in the MS, and the final output is robust to noise in individual trajectories.  These regions are statistical steady-states: the ensemble average of the RC states for each input are in the vicinity of their fixed points, with limited diffusion resulting from individual stochastic trajectories. The \textit{fading memory} property requires that the current RC state depends on the history of the input signal, with an increasing importance placed on more recent inputs.  Here, the final RC state depends more strongly on the steady state readout signal, where the SNR is higher and QS states can be more easily distinguished, aiding classification. At the same time, the RC state does not depend \textit{entirely} on its most recent input, making the classification result resilient to sudden changes in the input signal, for instance those due to qubit decay.

The final requirement of a RC is of some degree of \textit{nonlinearity} in its dynamics or output layer for nontrivial computation. For the linear RC, note that the classes approximately lie on a line in the MS. The linearity of {the $\bar{\Lambda} = 0$ RC enables its dynamics to be solved analytically:
$
\beta_j(t) =  \sum_k c_{jk}\int^t d\tau e^{(i\delta_k +\gamma/2)(\tau-t)}u(\tau)
$ 
where $\delta_k$ are the real parts of the eigenvalues of $\bm{W}_R$, and  $c_{jk}$ is the product of the projections of eigenvector $k$ onto node $j$ and the input $\bm{W}_I$.
 In the steady state, the different nodes $x_j$  are now effectively scaled and rotated copies of each other. The information capacity of classifying the final RC outputs is then no different from that of classifying the input signals themselves with a linear classifier.

By comparison, the role played by the Kerr nonlinearity to provide high fidelities for this task is evident from Fig.~\ref{fig:phasedyn}. The RC's nonlinearity `shears' the high-amplitude readout signals (associated with states $01$ and $10$) out of the line connecting the low-amplitude signals (states $00$ and $11$).  The measured RC quadratures are then linearly independent and not trivially related to each other or the input signal, in contrast to the linear RC.  The nonlinearity of the Kerr reservoir allows it to utilize its dimensionality, forming up to $K$ linearly independent outputs, rather than being bound by the dimension of the input signal.  Nonlinearity, together with the other dynamical RC properties satisfied by the Kerr RC, thus allow the four different classes of input signals and their resultant RC state distributions to be linearly separated via the output layer.  That the Kerr RC satisfies these properties is by no means guaranteed in general, but a result of the RC hyperparameters we have chosen.  In particular, it is strongly dependent on the relative strength of the nonlinearity $\bar{\Lambda}$ and the input signal scaling factor $\mu$, as discussed later in Sec.~\ref{subsec:RCdesign}. 

Finally, we discuss why the RC appears to require fewer training signals than an experimentally constructed MF to attain high-fidelity classification.   In Fig.~\ref{fig:phasedyn} we show the response to a \textit{single} input trajectory for each class of input signal in black. Even though these are unlabeled, it is clear what their ground truth initial quantum state was; an effective RC is able to integrate out the noise in the readout signals, enabling efficient separation of the input signals into the corresponding color-coded phase space distributions. These trajectories are information dense; each time point functions as a new point of training data.
Hyperplanes drawn to separate these trajectories are found to be differ only slightly from those shown on the plot, constructed using $Q=40$ trajectories, thus indicating that training from a small set of measurement signals can yield comparable performance to training using a larger training set.

This is in stark contrast to the MF approach, which  uses the noisy training data to construct a time dependent kernel, as opposed to static hyperplanes.  The MF needs to be able to separate signals at each point in time, and so a large number of training sets are needed to produce an estimate of the mean input at each time which is not dominated by noise. It is then reasonable to ask why these linear filters, which perform only a linear operation on the input signal, are able to perform the classification task with high fidelity, but a linear RC is not. The answer lies in the classification step: the continuous filtered signal $y^{(q)}(t)$ is mapped to a discrete class label $z$ by comparing which expected signal $y^{(q)}(t)$ is closest to, via a `distance' calculation that introduces the necessary nonlinearity. The RC approach is very different: the nonlinearity occurs in the dynamical signal processing, and the output classification step $\bm{y}=\bm{W}_o \bm{x}^{\bm{\phi}}$ is linear. The use of a nonlinear classification step would then supply the required nonlinearity and enable the linear RC to perform comparably to a nonlinear RC.

\subsection{Optimization of the Kerr Reservoir}
\label{subsec:RCdesign}


In the previous sections, we have investigated the ability of specific Kerr RC networks to perform a quantum measurement task.  We also demonstrated that this performance is not particularly dependent on that network structure by completing the same task with a set of random networks that share the same hyperparameters. We now explore the role of these hyperparameters in determining RC performance, presenting the dispersive system readout fidelity of random RC networks as {a function of $\gamma$, $\bar{\Lambda}$, and $\mu$ in Fig.~\ref{fig:hpvar}}.  

In Fig.~\ref{fig:hpvar}(a), we plot $\mathcal{F}$ for $10$ random RCs with $K=2,5,10$ (blue, green, and brown respectively), but all sharing the same hyperparameters. We vary \textit{only} the hyperparameter $\gamma$, which sets the rate at which the RC nodes evolve and thus the timescale over which it samples the input signal. This system response time is typically only considered for hardware RCs; in software approaches an RC conventionally evolves under an update that is equivalent to
$\gamma=1/\Delta t$ \cite{Jaeger2009}.
From Fig.~\ref{fig:hpvar}, we see that it is important for $\gamma$ to be approximately matched to the timescale of the input signal's evolution \cite{Canaday2018}; for the quantum readout task, the signal $\sqrt{\kappa}\expect{\hat{d} +\hat{d}^\dagger}_c$ evolves at rate $\kappa$. To understand this relationship, consider first a slow RC, with $\gamma\leq 0.1\kappa$: the RC then responds to the input signal averaged over a large window, 
$\beta_j\propto \int^t d\tau e^{\gamma/2(\tau-t)}u(\tau)$
and it is consequently more difficult to distinguish between different signals on the timescale over which the measurement is done.  This slow evolution furthermore results in very little displacement from the initial RC state $\bm{\beta}=0$ over this $10/\kappa$ measurement window.  Since the nonlinearity is $\propto \beta_j^3$, the low node amplitude also results in the RC being effectively linear.

There is a second timescale in the input signal: that of the noise $\xi(t)$, at the sampling rate of the quantum measurement $\Delta t$.  This defines an upper limit on $\gamma$ for good performance; it is advantageous for the RC to respond to the much slower underlying quantum signal rather than the rapid white noise, and so one should have $\gamma \ll {1}/{\Delta t}$. This allows the RC to average over some of the white noise, improving its performance.  As seen in Fig.~\ref{fig:hpvar}, $\gamma\sim 0.2-0.8\kappa$ is the roughly optimal range for the dispersive readout task, allowing the RC to still respond to signal dynamics while integrating out much of the readout noise.

\begin{figure}[t]
\captionsetup[figure]{labelformat=empty}
\includegraphics[width=0.47\textwidth]{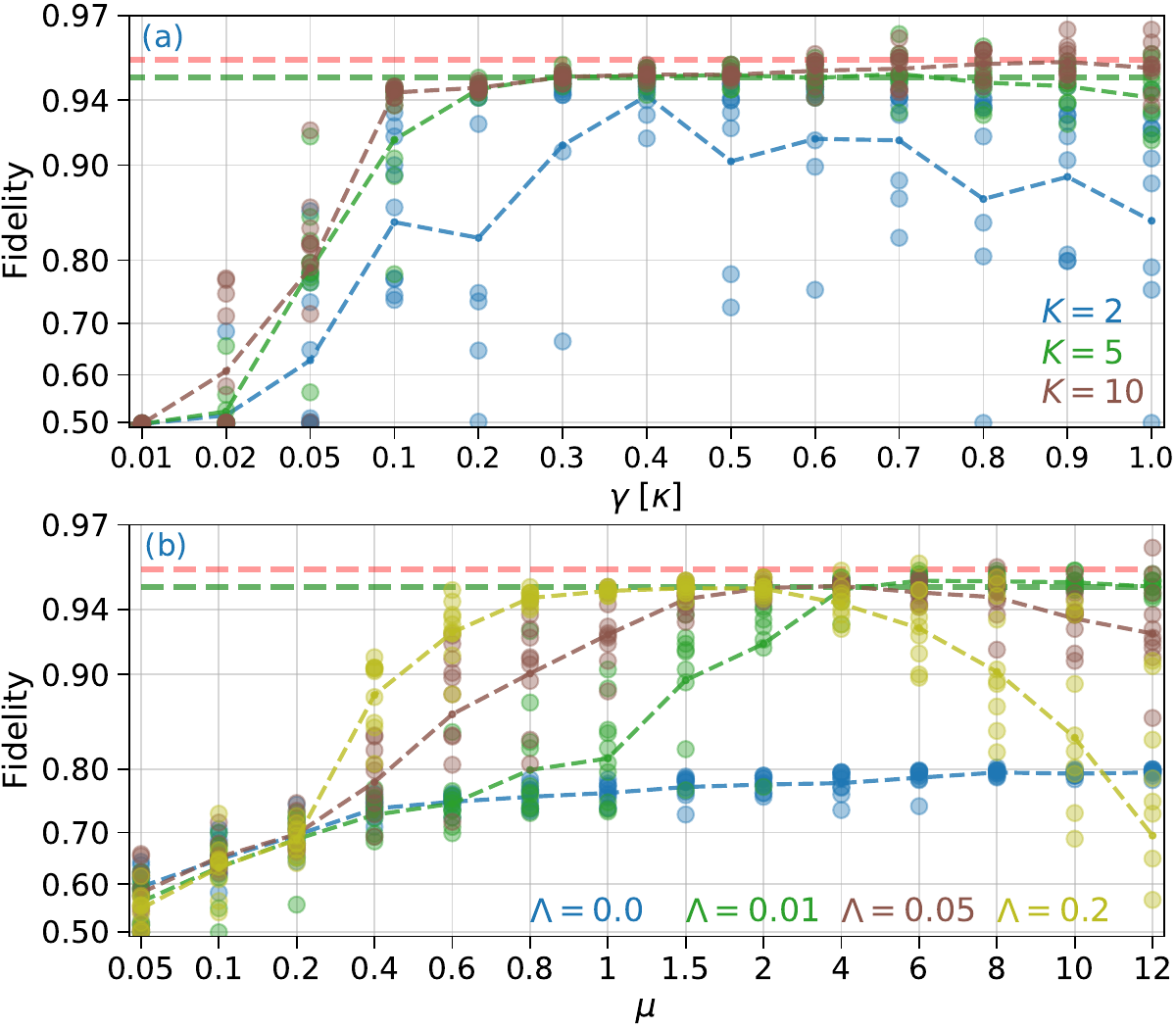}
\caption{Performance as a function of various hyperparameters. In (a) we vary the RC timescale $\gamma$ for RCs with $K=2$, $5$, and $10$ nodes, with $\mu=5$ and $\bar{\Lambda}=5\times10^{-2}$.  (b) shows fidelity as a function of input coupling strength $\mu$ and nonlinearity $\bar{\Lambda}$ for $K=5$ networks with $\gamma=0.5\kappa$. In all plots $\alpha=1.9$ and the RCs are trained with $Q=80$ measurements. $10$ different random networks are shown in (b), and $10$ for each $K$ in (a), with individual $\mathcal{F}$ in colored circles, and the average fidelity as a dashed line.  The fidelity of the boxcar and analytic MF are indicated in green and red dashed lines as in previous plots.
}
\label{fig:hpvar}
\end{figure}

Quite generally, with increasing $K$ the performance of the Kerr RCs becomes more robust to variation in hyperparameter values and to randomness in the structure ($\bm{W}_I$, $\bm{W}_R$, $\boldsymbol\Lambda$) of the RCs. For the optimal range of $\gamma\sim 0.2-0.8\kappa$, most of the RC networks have $\mathcal{F}\sim 0.95$ (effectively the theoretical limit).  Note however that out of the 30 simulated RCs for each $\gamma$, all the poorly performing networks in this range have $K=2$ (blue).  This can be understood as follows: since only two Kerr nodes are sufficient to perform the dispersive readout task, as $K$ increases so does the probability that some subset of the RC phase space forms a good network. If the other hyperparameters are within some reasonably optimal window, there is a $K$ for which the probability of finding a good sub-network is high enough that performance saturates and any Kerr RC achieves high fidelity. In fact, note that the $K=5$ and $K=10$ sets of networks show almost equivalent performance, indicating that as few as 5 nodes are sufficient for a random network to reliably complete the present readout task. Finally, we also find that the performance of larger reservoirs is more robust to $\gamma$ and other hyperparameter values outside the optimal range. 

In Fig.~\ref{fig:hpvar}(b) we consider the role of nonlinearity and input scaling, plotting $\mathcal{F}$ obtained using $10$ random RCs, while varying $\mu$ for various fixed values of $\bar{\Lambda}$.  The nonlinear term in Eq.~\eqref{eq:RCeom} scales as the node amplitude cubed, and as such the effective nonlinearity is related to both $\bar{\Lambda}$ and $\mu$.  To zeroth-order, $\boldsymbol\beta\propto\mu$, and so the effective nonlinearity in $\boldsymbol\beta$ and $\bm{x}^{\bm{\phi}}$ can be roughly quantified by $\bar{\Lambda}\mu^2$.  For a larger input strength, the field amplitude will be higher and thus a lower $\bar{\Lambda}$ is needed to produce a nonlinear term with the same relative weight. 

As we have seen in Sec.~\ref{subsec:phasedyn}, some nonlinearity is necessary for computation, so $\mathcal{F}$ initially increases as either $\bar{\Lambda}$ or $\mu$ is increased. {However, the fidelity ultimately reaches a maximum before decreasing again, as we encounter} an upper limit to the nonlinearity:  if it is too strong, the fixed points of the RC network for a given steady state drive become less stable.  The dynamics of the RC network will generically exhibit large oscillations and not settle near their steady-states over the measurement timescale.  This is exacerbated by the noise in the input signal; as these fixed points in phase space are less attractive, the strong white noise is able to generate larger excursions in phase space.  Thus, as seen in Fig~\ref{fig:hpvar}(b), $\mathcal{F}$ falls off sharply for larger $\bar{\Lambda}$ for $\mu$ above some upper limit.  Indeed, we find that $\sqrt{\bar{\Lambda}}\mu$ should be in the range of  $ 0.5-1$ for optimal performance, a trend we verified for additional $\bar{\Lambda}$ which are not shown.  This is precisely the regime for a single Kerr oscillator where the nonlinearity begins to significantly influence its dynamics and steady-state. {RC intuition suggests that dynamics should be affected but not dominated by the nonlinearity; the observation of an optimal nonlinearity strength for a Kerr network appears consistent with this understanding.}  It should be noted that for this plot, we have chosen $\Lambda_j=\bar{\Lambda}$ to make the $\bar{\Lambda}-\mu$ relationship more clear, but the results are qualitatively unchanged when randomness in the nonlinearity is reintroduced as well.

The final hyperparameter $\alpha$ is the largest singular value of the Kerr network connectivity matrix $\bm{ W}_R$ and sets the strength of the node-node coupling.  In reservoir computing literature, it is commonly stated that for an echo-state network (with a hyperbolic tangent nonlinearity), if the spectral radius of $\bm{ W}_R$ is much larger than 1,
the RC steady state is not guaranteed to be stable and limit-cycle dynamics can emerge \cite{Jaeger2009, Larger2017, Canaday2018}. 
A spectral radius close to this limit (the so-called `edge of stability') is often found to be optimal for tasks requiring significant memory (i.e.~where previous states of the input are important)~\cite{Dambre2012,Canaday2018}.  For the Kerr network RC of Eq.~\eqref{eq:RCeom}, the coupling matrix is symmetric so $\alpha$ is also the spectral radius of $\bm{ W}_R$.  Since the network has an explicit decay term, the linear network is stable for  $\alpha<2$. When the nonlinearity is included, numerically we find stable steady states and thus the fading memory property are always present for $\alpha<2$ for the nonlinearity parameters we consider.  In agreement with other works, we have found $\alpha\sim 1.5-2$ results in optimal performance, and thus have chosen to present results with $\alpha=1.9$.

Overall, there is generally a broad range of Kerr RC hyperparameters resulting in high fidelity classification, which we can summarize as $\gamma \lesssim \kappa$, $\sqrt{\bar{\Lambda}}\mu \sim 0.5-1$, and $\alpha\leq 2$.  This performance is independent of specific network structure, and is robust to moderate disorder in these structural parameters.

\section{Quantum Information Applications}
\label{sec:Qinfo}
{
In this final section, we demonstrate a pair of relevant quantum information applications that can be implemented in this same reservoir computing system: two-qubit state tomography and continuous qubit parity monitoring.  In both cases the description of Fig.~\ref{fig:scheme} applies: a Kerr network reservoir continually processes the measurement current from a joint dispersive readout system.  This is not intended to be an exhaustive survey of potential applications, but to emphasize the ease of generalizing our reservoir processing approach.  
}

\subsection{Multi-qubit tomography}
{
To this point, we have only evaluated the ability of the RC to classify measurement currents from quantum systems prepared in computational basis eigenstates.  However, a reservoir processor trained using only these states  $\ket{\psi(0)} =\ket{z}$ can measure \textit{arbitrary} joint qubit states with high fidelity.  To be specific, when the target quantum system is interrogated by driving the readout cavity, backaction rapidly causes the joint qubit state to collapse to one of the measurement basis eigenstates, with probability $\simeq|c_z(0)|^2$.  The RC will then faithfully return the current state of the target quantum system $\ket{\psi(\tau_m)}$. If this quantum state is repeatedly prepared and measured, the distribution of RC outputs will thus agree with that of the underlying state.  Furthermore, since the measured cavity quadrature is a nonlinear function of the multi-qubit operator $\hat{\chi} = \sum_j \chi_j\hat{\sigma}_{z, j}$ (Appendix \ref{app:readout}), one can preform full tomography on the two-qubit density matrix by simply preceding the measurement with a set of single-qubit rotations \cite{Filipp2009}.

\begin{figure}[t]
\captionsetup[figure]{labelformat=empty}
\includegraphics[width=0.48\textwidth]{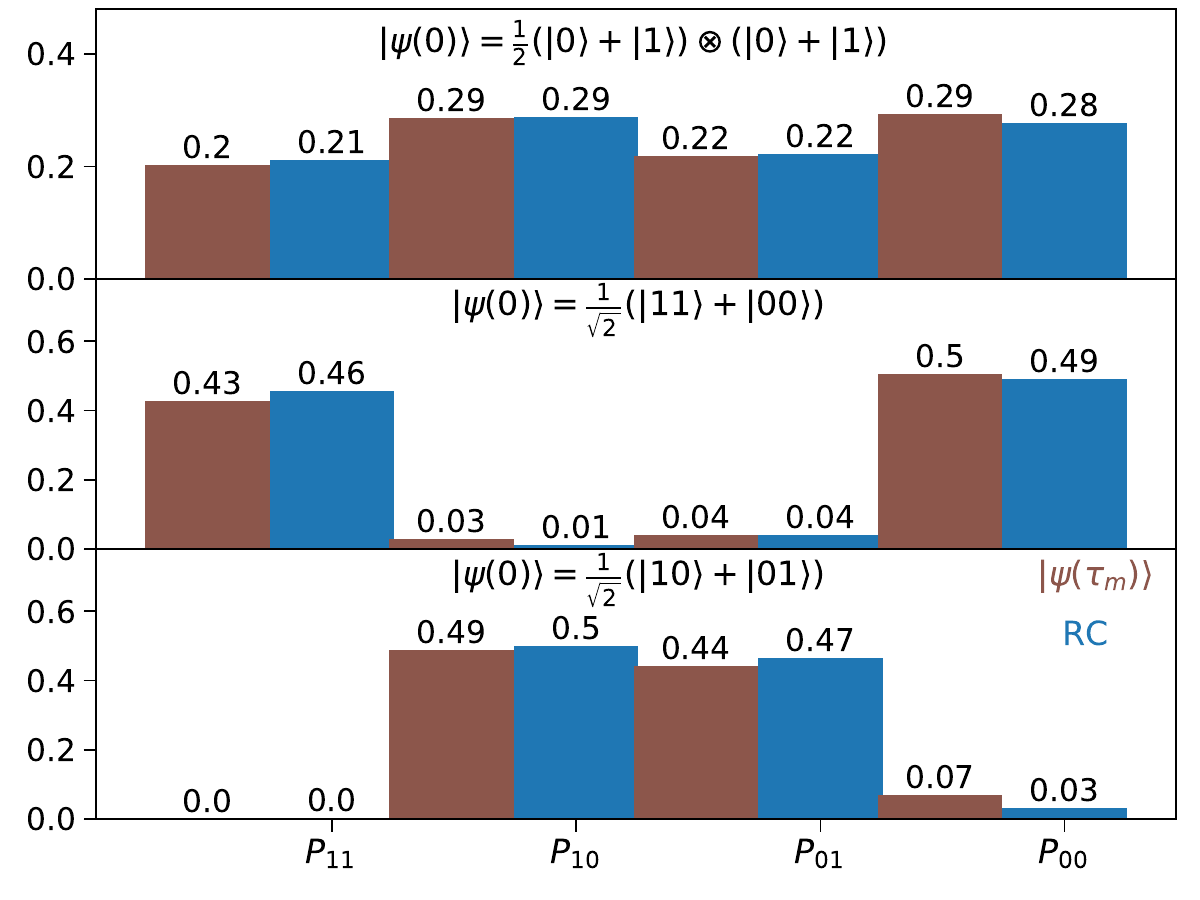}
\caption{Tomography demonstration:  the dispersive system is prepared in the indicated initial state and measured $800$ times.  Measurement causes these superposition states to collapse; the true distribution in the computational basis at $\tau_m=7.5/\kappa$ is indicated in red.  In blue we plot the distribution of classification outputs for the RC of  Figs.~\protect\ref{fig:claDyn}-\protect\ref{fig:claFD}, and in particular the same $\bm{W}_o$, trained on $Q=40$ computational basis initial states.
}
\label{fig:tomography}

\end{figure}

To demonstrate this capability,  in Fig.~\ref{fig:tomography} we compare the RC output with the true quantum state at the measurement time, for the dispersive system initialized in the indicated product or Bell states.  The reservoir is that of Fig.~\ref{fig:claDyn}-\ref{fig:claFD}, and in particular has the same $\bm{W}_o$: training was again done by simply preparing the qubits in each computational basis state 10 times, measuring for $10/\kappa$, and using that initial state as the target.   Even though the qubit state can jump during this readout, training is still effective, and this approach should be robust to preparation errors as well.   The quantum system was then initialized in each of the test superposition states 800 times, and the time-dependent RC output $P_z(t)$ was compared against the current quantum state $|c_z(t)|^2$.  Despite this simple training with only easily-accessible initial computational basis state labels, the RC is highly successful at producing this dynamical quantum variable, returning the current state $\ket{\psi(t)}$ (${\rm max }_z\{P_z(t)\} ={\rm max}_z\{|c_z(t)|^2\}$) with 98\% fidelity across all test states.

Fig.~\ref{fig:tomography} compares the distribution of RC outputs with the average quantum state at $\tau_m$; it is clear the RC accurately determines the underlying quantum distribution for the three states tested. 
For simplicity we have chosen a set of states where all the unique density matrix elements are diagonal in the measurement basis, so this single set of measurements is sufficient to distinguish states. As described in Ref~\onlinecite{Filipp2009}, full tomography on arbitrary states can be done by repeating this process after applying single qubit rotations; this reservoir processor can thus be an effective tool for general tomography in additionto computational basis measurement.
}

\subsection{Joint parity monitoring}

{
This reservoir processing approach is not limited to determining qubit states from measurement currents - generally, one can train an RC to return arbitrary dynamical observables given an appropriate measurement record.  In this final example we describe the operation of the $K=5$ Kerr reservoir of Fig.~\ref{fig:claDyn}-\ref{fig:claFD} to measure multi-qubit parity: $\expect{\hat{\sigma}_{z, 1}\hat{\sigma}_{z, 2}}$.  Recall that for the dispersive readout quantum system model described previously, the RC was able to learn the quantum state and output $|c_z(t)|^2$ with very high fidelity: thus, one can trivially modify the output layer to instead return the expected parity
via $\bm{W}_o \to (1, -1, -1, 1) \cdot \bm{W}_o$. 

We instead consider a more interesting and relevant task for quantum information applications by modifying the dispersive readout system such that the observed quadrature does not distinguish between states in a given parity subspace, setting $\chi_1 = -\chi_2 =\kappa$, $\varepsilon_0=2i\kappa$, and leaving all other parameters unchanged.  In this situation, the measured quadrature will differ only for states of different parity, and be the same for states in the even $\{\ket{11},\, \ket{00}\}$ or odd parity sub-spaces $\{\ket{11},\, \ket{00}\}$ (Eq.~\eqref{eq:alphasol}). As a result, when the readout cavity is measured there is no backaction on qubit states in a given parity subspace: this allows one to generate and maintain superposition states, or manipulate the joint qubit state within a parity subspace, a requirement for many quantum error correction protocols.  This specific parity readout system has been explored both for Bell state generation and error syndrome monitoring \cite{Lalumiere2010, Riste2013, Nigg2013}.  In particular, this is a common error syndrome in quantum error correction, where the detection of a change in parity between two qubits is an unambiguous indicator that an error has occurred \cite{Nigg2013, NielsenChuang2010}.

}

\begin{figure}[t]
\captionsetup[figure]{labelformat=empty}
\includegraphics[width=0.48\textwidth]{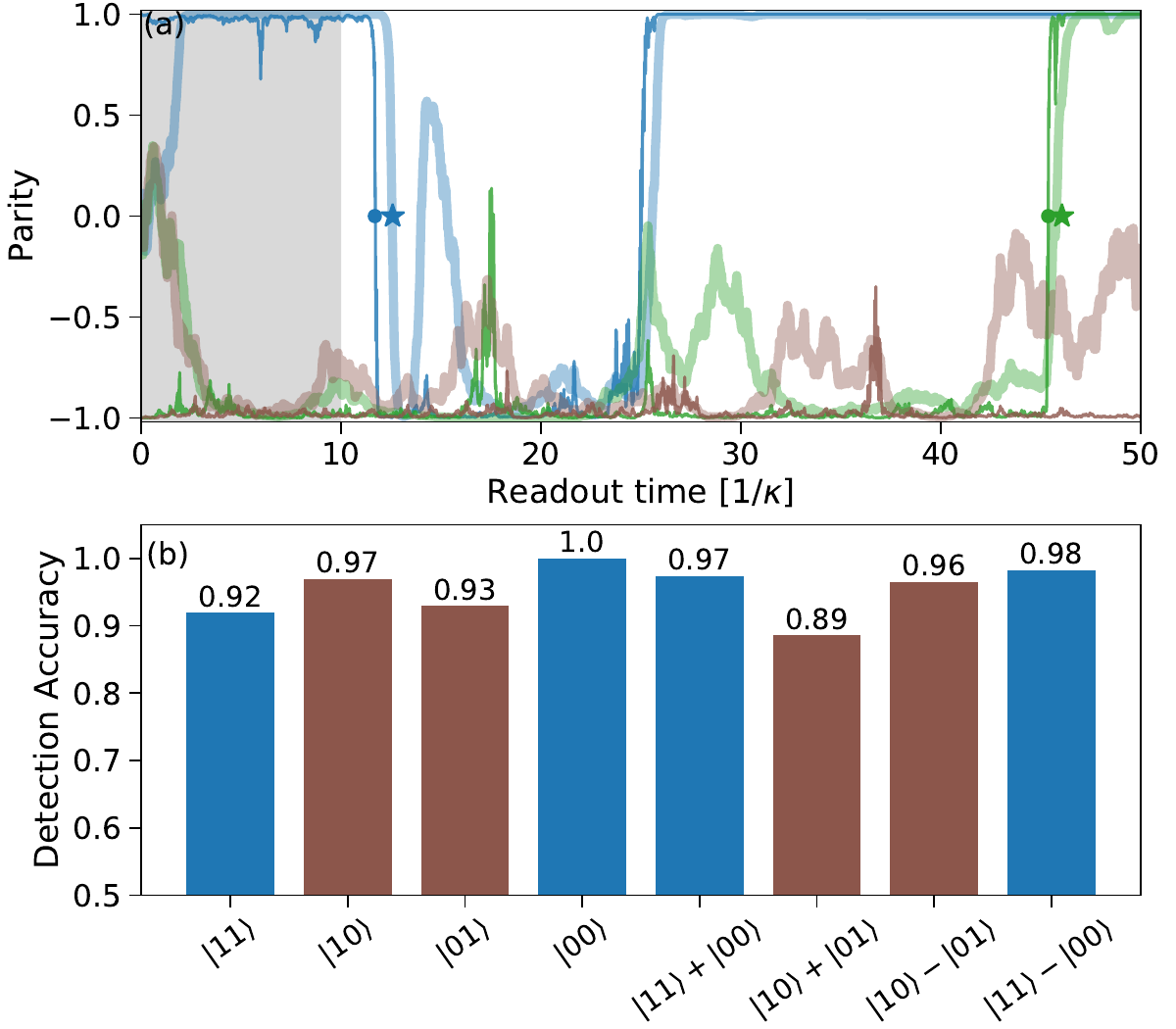}
\caption{(a) Ground truth parity trajectories in dark thin lines and corresponding RC parity estimate in light thick lines of the same colours.  The times at which the qubit parity jumps are indicated with circles, the resultant RC parity change is denoted with a star.  The RC of Figs.~\protect\ref{fig:claDyn}-\protect\ref{fig:claFD}
is again used, trained with $Q=40$ measurements over the $10/\kappa$ window shaded in grey.
(b) Fraction of trajectories for which parity monitoring was successful over the entire $50/\kappa$ monitoring window for each initial state.  Failures occur when either the RC predicts a parity change where none occurred, or fails to detect a parity change.  Odd and even parity initial states are colored red and blue respectively.
}
\label{fig:parity}
\end{figure}

{
We operate this system as previously, interrogating the measurement cavity and inputting the resultant measurement current to the RC; this task is thus also described via Fig.~\ref{fig:scheme}, with the output now $\expect{\hat{\sigma}_{z, 1}\hat{\sigma}_{z, 2}}$ instead of $P_j$.  For training, the quantum system is again prepared in each of the  computational basis states $Q=40$ times and a measurement is performed for $10/\kappa$, with the target being the initial parity, $\pm1$ for $\ket{\psi(0)}=\ket{11/00}, \, \ket{10/01}$ respectively.  For the parameters chosen, the probability of a parity jump occurring during this window is small due to the lack of measurement backaction; thus this simple training procedure allows the RC to learn to return the current multiqubit parity.  This is demonstrated in Fig.~\ref{fig:parity}(a), where we test the ability of the RC to continuously return the parity as the quantum system is measured over a much longer $50/\kappa$ window.  For sample readout trajectories, we have plotted both the evolution of the parity (ground truth obtained from the SME) as well as the RC output.  It is seen that the RC output follows the true parity closely, and in particular, quickly switches after the parity jumps due to qubit decay processes.  Over 800 test measurements (each of duration $50/\kappa$), the RC parity is correct at $93\%$ of times.  This is limited by the finite response time of the reservoir: on average the RC parity will flip $2.3/\kappa\sim 1/\gamma$ after that of the quantum system. 

For both error syndrome monitoring and Bell state generation, it is only necessary to detect these parity jumps, rather than exactly reproduce the evolution of the parity
. In Fig.~\ref{fig:parity}(b) we indicate the times of parity jumps in the quantum system and RC output with circles and stars respectively.  It is clear that for the examples shown, the RC accurately detects when a parity jump occurs, and does not predict one where there is no jump.  In Fig.~\ref{fig:parity}(b) we test the ability of the RC to detect parity jumps:  the quantum system is initialized in the 8 indicated states (of definite parity), and measured 100 times each over a duration of $50/\kappa$.  Plotted is the fraction of measurements in which the RC successfully predicted a parity jump after one occurred in the quantum system, or did not predict a parity jump if none occurred.  Overall, parity change detection accuracy was $95.5\%$, indicating that this RC could be a powerful tool for monitoring error syndromes or tracking the evolution of general observables.

}

\section{Conclusions}

In this work, we propose a hardware reservoir computer to facilitate quantum readout. We describe how a small network of Kerr nonlinear oscillators can be implemented and integrated with superconducting circuit or conventional optical quantum systems.  These RCs can reliably perform the nontrivial task of joint dispersive quantum measurement with comparable high fidelity and orders of magnitude less training overhead than conventional filtering approaches.
{
This processor can be readily applied to important quantum information tasks such as multi-qubit tomography and the continuous monitoring of observables such as parity, with similar fidelity and simplicity of calibration. This approach offers significant efficiency improvements across several dimensions: (i) hardware resources via compatibility with existing quantum platforms, (ii) computational resources via ease of training, and (iii) latency and data overhead via edge-computing in a physical reservoir.
}
Through a first-principles consideration of the system dynamics, we explore the features and properties of the Kerr oscillator network that enable this performance and make it an attractive platform for reservoir computing more generally.  We additionally develop an intuitive phase-space picture which provides insight into how RCs process information.  Here we have considered only a classical RC which is not entangled with the quantum system it measures, but extend our approach to the quantum domain reservoir computer in a separate publication. We hope that this work helps support the development and integration of hardware-based reservoir computing approaches to quantum information processing platforms, where we believe they can be of significant benefit.

\section{Acknowledgements}

We thank Daniel J. Gauthier, Michael Hatridge, Peter L. McMahon, Shyam Shankar, and Nikolas Tezak for discussions.  This work is supported by NSERC and AFOSR Grant No. FA9550-20-1-0177.

\appendix

\section{Table of notations}

\begin{table}
\caption{
{
Summary of symbols and parameters in this paper}
}
\begin{tabularx}{.48\textwidth}{@{}p{0.04\textwidth}X@{}}
\toprule
        \multicolumn{2}{l}
        {Measured quantum system}  \\
    \hline
     $\hat{d}$ & cavity field operator \\
     $\hat{\boldsymbol\sigma}_j$ & qubit Pauli operator \\
     $J^{(q)}$ & cavity measurement current, sample $q$ \\
    $\kappa$ & readout cavity decay rate \\
    $\omega_d$ & frequency of cavity drive, measurement current \\
    $\Delta_c$ & readout cavity - drive detuning: $\omega_c-\omega_d$ \\
    ${\Delta}_{q, j}$ & qubit $j$ - drive detuning: $\omega_{q, j}-\omega_d$ \\
    ${\delta}_{j}$ & qubit $j$ - cavity detuining: $\omega_{q, j}-\omega_c$ \\
    $g_j$ & qubit $j$ - cavity direct coupling \\
    $\chi_j$ & qubit $j$ dispersive shit: $g_j^2/\delta_j$ \\
    $\gamma_h$ & qubit decay rate (non-Purcell) \\
    $J_{12}$ & qubit-qubit coupling via cavity \\
    $\tau_m$ & duration of measurement signal \\
    \hline
    \multicolumn{2}{l}
    {Kerr network reservoir} \\
    \hline    
     $\hat{b}_k$ & Kerr oscillator $k$ field operator \\
     ${\beta}_k$ & Kerr oscillator $k$ field amplitude \\
     ${\bm u}$ & input signals to Kerr network at $\omega_d$ \\
     $\Delta_k$ & Kerr oscillator $k$ - input detuning: $\omega_k-\omega_d$ \\
     $\lambda_k$ &Kerr oscillator $k$ nonlinearity\\
    $g_{kl}$ & linear coupling between Kerr oscillators $k$ and $l$ \\
    $\varepsilon_{km}$ & input $m$ -  Kerr oscillator $k$ coupling strength \\
    $\gamma$ & Kerr oscillator decay rate, network evolution rate \\
    \hline
    \multicolumn{2}{l}
    {Reservoir computer}  \\    \hline   
    $\bm{x^\phi}$ & measured reservoir quadratures:
    $(\beta_k e^{-\phi_k}+\beta_k^* e^{i\phi_k})/\sqrt{2}$
    \\
    $\bm{W}_{R}$ & dimensionless connectivity matrix:
    $2(\Delta_k \delta_{k,l} +g_{kl})/\gamma$
    \\
    $\bm{W}_I$ & dimensionless input layer:
    $2\tilde{\varepsilon}_{kl} /\gamma$
    \\
    $\bm{\Lambda}$ & dimensionless node nonlinearity $2\tilde{\lambda}_k/\gamma$ \\
    $K$ & number of Kerr oscillators \\
    $\mu$ & input coupling strength: $W_{I,kl} \in [-\mu, \mu]$ \\
    $\bar{\Lambda}$ & average node nonlinearity: $\Lambda_k \in [0, 2\bar{\Lambda}]$ \\
    $\alpha$ & spectral radius of connectivity matrix: $\lambda_{max}(\bm{W}_R)$ \\
    $\bm{C}(\bm{\phi})$ & matrix of node measurement angles \\
    $\bm{W}_o$ & trained output layer \\
    $Q$ & number of training samples \\
    $\bm{y}^\star$ & $C$-dimensional target output \\
    $\bm{y}$ & RC output: $\bm{W}_o\bm{x}^{\bm{\phi}}$ \\
    $P_z$ & RC output converted to probability $e^{{ y}_z(t)}/\sum_k e^{{ y}_k(t)}$ \\
    $C_{\mathcal{F}}$ & fraction of test signals classified correctly \\
    $\mathcal{F}$ & classification fidelity: $ {\rm max}_t \{C_{\mathcal{F}}(t)\}$ \\
\end{tabularx}
\end{table}
}
\section{Joint Dispersive Measurement}
\label{app:readout}

The unconditional evolution of the multi-qubit measurement system under either model is described by the master equation (ME) \cite{Filipp2009, Hutchison2009, Lalumiere2010}:
\begin{align}
\dot{\hat{\rho}} =\mathcal{L}\hat{\rho}
= -i[\hat{\mathcal{H}}_{S}, \hat{\rho}]  
+\gamma_h\sum_j \mathcal{D}[\hat{\sigma}_{-,j}]\hat{\rho}
+\kappa \mathcal{D}[\hat{d}]\hat{\rho}
\label{eq:sysme}  
\end{align}
where $\kappa$ and $\gamma_h$ describe cavity loss and qubit decay through coupling to the external environment. $\hat{\mathcal{H}}_{S} = \{\hat{\mathcal{H}}_{JC}, \hat{\mathcal{H}}_D\}$: for $\hat{\mathcal{H}}_{S} = \hat{\mathcal{H}}_D$ the ME acquires an additional term to account for correlated qubit decay via the cavity: $\mathcal{L}\to \mathcal{L} +\kappa\mathcal{D}[\sum_j\frac{g_j}{\delta_j}\hat{\sigma}_{-,j}]$, although it is $O((\frac{g_j}{\delta_j})^2)$ and thus weaker than other rates considered.  We could also include pure dephasing as well but consider a regime where this environmental dephasing much weaker than dephasing  induced by measurement.

The utility behind this  dispersive measurement system is that it allows one to perform a near-QND measurement of all the qubits simultaneously. We recall first the basic description of the measurement process from Ref.~\onlinecite{Gambetta2008}, based on the pointer-state formalism.   With the cavity initially in the vacuum and a constant drive applied at $t=0$, under $\hat{\mathcal{H}}_{S} = \hat{\mathcal{H}}_D$  there is a cavity coherent state associated with each joint qubit state (i.e.~$\hat{\rho}\propto \sum_{z, z'}p(t)_{z, z'}\ketbra{\alpha_{ z}(t),{z}}{\alpha_{ z'}(t),{z'}}$. The coherent state amplitudes evolve as
\be
\frac{d}{dt}\alpha_{ z}(t) = -i\epsilon_0 - \big(i(\Delta_c +\expect{\hat{\chi}}_{ z})+ \frac{\kappa}{2}\big)\alpha_{ z}(t)
\label{eq:alphade}
\ee
where we have defined the  qubit operator $\hat{\chi} = \sum_j \chi_j\hat{\sigma}_{z, j}$.  If $\chi_j$ are distinct then for each of the $2^{N_q}$ $\{\ket{{z}}\}$ qubit states, $\hat{\chi}$ has a unique value and a different coherent state is generated in the measurement cavity.  The corresponding steady-states are
\be
\alpha_{{ z}, ss} = -\epsilon_0\frac{\Delta_c +\expect{\hat{\chi}}_{ z}
+ i\frac{\kappa}{2} }{(\Delta_c +\expect{\hat{\chi}}_{ z} )^2 + (\frac{\kappa}{2})^2 }
\label{eq:alphasol}
\ee
And the measurement of this cavity field results in an effective measurement of the joint qubit state, i.e.~all qubits in the $z$-basis simultaneously.  This measurement becomes QND if $\hat{\chi}$ commutes with all the operators in $\mathcal{L}$, such that joint qubit-eigenstates are preserved by the measurement process.  This can be achieved in practice in the dispersive regime if $\kappa$ is larger than $J_{jk}$ and the qubit decay rates, such that these terms can be ignored on the timescale over which measurement proceeds.  Even away from the dispersive regime, where the system is better described by $\hat{\mathcal{H}}_{JC}$, the state of the cavity field still contains information about the joint qubit state, and a sufficiently rapid cavity measurement can be used to learn the initial joint qubit state $\ket{\psi(0)}$.

By continuously recording the $X$-quadrature of the field radiating from the measurement cavity, one obtains the homodyne measurement current of Eq.~\eqref{eq:Jhom}.  This continuous measurement provides information about the current state of the QS, $\hat{\rho}(t)$, and the evolution of an observers knowledge of the QS, conditioned on $J(t)$ is found by including the  measurement superoperator to Eq.~\eqref{eq:sysme}, resulting in the SME of Eq.~\eqref{eq:syssme}.  Recall that  $\expect{\xi(t)}=0$, and so by taking the ensemble average of Eq.~\eqref{eq:syssme} we recover Eq.~\eqref{eq:sysme} and $\expect{J(t)}=\sqrt{\kappa}\expect{d+d^\dagger}$:  density matrix evolution and operator expecation values converge to their unconditional results.

Equation \eqref{eq:Jhom} describes a measurement current and corresponding noise which is continuous; both due to finite sampling times in real measurements and for our numerical simulations, we the homodyne current is actually sampled at discrete times $t_n= n\Delta t$.  As a consequence, $\xi(t)\to \xi(t_n)$ in both Eq.~\eqref{eq:syssme} and \eqref{eq:Jhom}:
\be
\xi(t_n)= \mathcal{N}_{n}(0, 1)/\sqrt{\Delta t}
\ee
where $ \mathcal{N}_{n}(0, 1)$ are samples drawn from the normal distribution with zero mean and unit variance.  This converges to white noise in the continuum limit $\Delta t \to 0$, with ${\expect{\xi(t)\xi(t')}} \to \delta(t-t')$.  The integrated noise power is importantly independent of $\Delta t$: $\int^t d\tau d\tau' {\expect{\xi(\tau)\xi(\tau'))}} = t$, so the SNR of the measurement current remains fixed for any sampling rate.

The noisy continuous signal described in Eq.~\eqref{eq:Jhom} is what is seen by any classical system interacting with the readout cavity output field \cite{Jacobs2006}.  When we take $J(t)$ as the input to an RC,  there is no entanglement between QS being measured and the classical
RC processing this measurement record.  In this work we assume each Kerr oscillator has sufficiently large field amplitudes to be in a classical regime.  Even without a standard homodyne measurement set-up, the signal is measured through its interaction with room temperature
electronics or the RC itself.  If the RC and QS are built on different hardware platforms, or separated from the QS by amplification and isolation stages, this description is necessary.


\section{Reservoir Computer Training}
\label{app:Train}

Training amounts to choosing an optimal set of phase angles to measure each oscillator and linear weights to apply to to each resultant node.  For a $K$ node network and $C$-dimensional target output, these are encoded in the $C \times K$  output matrix $\bm{W}_o$ and the $K$ element vector $\bm{\phi}=(\phi_1\ldots \phi_K)$ of measurement angles, for a total of $(C+1)\dot K$ parameters to optimize.  To train an RC, a set of labelled training data $\{ \bm{ u}^{(q)}(t), \bm{y}^{\star (q)}(t)\}$, consisting of $Q$ (generally multi-dimensional) input signals $\bm{U}=\bm{ u}^{(q)}(t)$ and their respective target outputs ${\bm{Y}}^{\star} = {\bm{y}}^{\star (q)}(t)$ is constructed.  The training data is fed into the RC, producing the dynamical response $\bm{ \beta}^{(q)} (t)$, which we decompose into independent quadratures $\bm{x}^{(q)} (t)$.  Importantly, both quadratures should be measured during training, in order to optimize the measurement angle. A loss function $\mathcal{L}(\bm{W}_o,\bm{\phi})$ is then constructed from the training RC trajectories and target outputs, which depends on the difference between the RC and target output, and thus the output matrix and set of measurement angles. By minimizing the loss function with respect to $\bm{W}_o$ and $\bm{\phi}$, one hopes to have the RC output $\bm{y}^{(q)}(t)={W}_o\bm{C}(\bm{\phi})  \bm{x}^{(q)}(t)\to {\bm{ y}}^{\star (q)}(t)$ and thus reproduce the desired target function. 

The specific task we consider in this work is the retrodiction of the initial state that produced an observed measurement record. We demand that the computation returns the probability that the input is a homodyne record from a QS with $\ket{\psi(0)}=\ket{z}$.  The training data labels are similarly probabilities for each input signal,  ${\bm{P}}^{\star (q)}$, which in practice is just 1 for the ground truth class and zero for the others. The cross entropy loss function is then
\be	
\mathcal{L}_x( \bm{W}_o, \bm{\phi}) = -\frac{1}{QN}\sum_{q, t_n} {\bm{P}}^{\star (q)}(t_n) \cdot \log( \bm{P}^{(q)} (t_n))
\label{eq:mxent}
\ee
where  $\bm{P}^{(q)} (t_n)$ is found from ${\bm{y}}^{(q)}(t)$ via Eq.~\eqref{eq:Pout}.

Throughout this work, training is done by minimizing $\mathcal{L}_x$  of Eq.~\eqref{eq:mxent} with an added regularization term through gradient descent. We consider small $K=2-10$  RCs and small training sets so this loss function minimization is a computationally easy task.   In our physical RC framework, we envision this training is done using external software to the RC. The RC node trajectories are measured during training, and $\{ \bm{x}^{(q)}(t), {\bm{ y}}^{\star (q)}(t)\}$ are used to minimize $\mathcal{L}_x$ to compute $\bm{W}_o$ and $\bm{\phi}$ for the task at hand.  $\bm{\phi}$ then sets the measurement angles for subsequent RC processing and $\bm{ W}_o$ maps these measurements to the RC output.  The RC output is the linear combination of measured node quadratures $\bm{y}^{(q)}(t)$, which reflects the probability of the input measurement record being associated with each state.  The softmax function is applied  in software and is only necessary to construct $\mathcal{L}_x$ during training; linear classification using $\bm{y}^{(q)}(t)$ or $\bm{P}^{(q)}(t)$ is equivalent.

\bibliography{paperrefs}
\end{document}